\documentclass[fleqn,usenatbib]{mnras}

\usepackage{newtxtext,newtxmath}
\usepackage{xcolor, ulem}

\usepackage[T1]{fontenc}

\DeclareRobustCommand{\VAN}[3]{#2}
\let\VANthebibliography\thebibliography
\def\thebibliography{\DeclareRobustCommand{\VAN}[3]{##3}\VANthebibliography}

\usepackage{graphicx}	
\usepackage{amsmath}	
\usepackage{pdflscape}

\usepackage{xurl}

\title[JWST MIRI LRS L and T Dwarf Library]{Silicate Clouds, SiO, and Molecular Trends across the L–T Sequence: A \textit{JWST}/MIRI LRS Spectral Library of Brown Dwarfs}

\author[S. Barber et al.]{
Sam Barber$^{1}$\thanks{E-mail: sbarbe24@uwo.ca},
Stanimir Metchev$^{1,2}$,
Genaro Suárez$^{3}$,
Samantha Lambier$^{1,2}$,
Paulo A. Miles-Páez$^{4}$,
\newauthor
\'Etienne Artigau$^{5,6}$,
Beth Biller$^{7,8}$,
Adam J. Burgasser$^{9}$,
Jacqueline Faherty$^{3}$,
Gregory Mace$^{10}$,
\newauthor
Caroline V. Morley$^{10}$,
Johanna M. Vos$^{11}$
\\
$^{1}$Department of Physics and Astronomy, The University of Western Ontario, 1151 Richmond St, London, Ontario N6A 3K7, Canada\\
$^{2}$Institute for Earth and Space Exploration, The University of Western Ontario, 1151 Richmond St, London, Ontario N6A 3K7, Canada\\
$^{3}$Department of Astrophysics, American Museum of Natural History, Central Park West at 79th Street, New York 10024, USA\\ 
$^{4}$Centro de Astrobiología, CSIC-INTA, Camino Bajo del Castillo s/n, E-28692 Villanueva de la Cañada, Madrid, Spain\\
$^{5}$Trottier Institute for Research on Exoplanets and Département de Physique, Université de Montréal, 1375 Avenue Thérèse-Lavoie-Roux,\\
Montréal, Quebec H2V 0B3, Canada\\
$^{6}$Observatoire du Mont-Mégantic, Université de Montréal, Montréal, Quebec H3C 3J7, Canada\\
$^{7}$Institute for Astronomy, University of Edinburgh, Royal Observatory, Edinburgh EH9 3HJ, UK\\
$^{8}$Centre for Exoplanet Science, University of Edinburgh, Edinburgh EH8 9YL, UK\\
$^{9}$Department of Astronomy \& Astrophysics, University of California San Diego, La Jolla, California 92093, USA\\
$^{10}$Department of Astronomy, The University of Texas at Austin, Austin, Texas 78712, USA\\
$^{11}$School of Physics, Trinity College Dublin, The University of Dublin, Dublin, Ireland 
}

\date{Accepted XXX. Received YYY; in original form ZZZ}

\pubyear{2025}

\begin{document}
\label{firstpage}
\pagerange{\pageref{firstpage}--\pageref{lastpage}}
\maketitle

\begin{abstract}
We present the results of a survey program with \textit{JWST}'s MIRI LRS that obtained 5--14\,$\mu$m R$\sim$100 spectra of twenty-three L0--T6 dwarfs. This spectral type range spans the formation, growth, and sedimentation of silicate condensate clouds, as well as the appearance of methane and ammonia in the atmospheres of brown dwarfs. The initial onset of silicate absorption at spectral type L0 as well as the evolution of this feature with spectral type are both examined in unprecedented detail. We detect 7.5--9.5\,$\umu$m gas-phase SiO absorption for the first time in dwarf stars, with detections in M5.5--L2.5 dwarfs from combined \textit{Spitzer} IRS and \textit{JWST} spectra. This feature has previously been detected in K and M giants, with stronger absorption in these giants than in higher-gravity M dwarfs. In contrast, low-gravity L dwarfs show weaker SiO absorption than field-gravity L dwarfs. We also confirm water, ammonia, and methane absorption trends with spectral type previously seen in \textit{Spitzer} IRS observations. We report the possible detection of CS$_2$ in a T3.5 dwarf, although confirmation would require higher-dispersion spectroscopy. We find two newly resolved binaries, confirm two near-equal flux unresolved candidate binaries, and identify two additional possible non-equal flux binaries. After accounting for contamination by resolved and candidate binaries, we find that single field dwarfs are $\sim$0.15--0.20 mag underluminous compared to existing empirical relations. The presented spectral library offers a spectrophotometrically calibrated basis for future interpretation of \textit{JWST} MIRI spectra of brown dwarfs or of directly imaged or transiting exoplanets.
\end{abstract}

\begin{keywords}
stars: atmospheres $-$ binaries: general $-$ brown dwarfs $-$ infrared: stars
\end{keywords}



\section{Introduction}
\label{sec:intro}

Brown dwarfs show a diversity in spectra in the mid-infrared (mid-IR; 5--14\,$\umu$m). Water absorption is ubiquitous in this wavelength range, while ammonia and methane absorption are observed starting at the L-T spectral type transition \citep[e.g.,][]{Cushing2006, Suarez2022}.

Amidst these molecular bands, the mid-IR also samples absorption due to silicate condensates in the atmospheres of brown dwarfs. Using the \textit{Spitzer} Infrared Spectrograph \citep[IRS,][]{Houck2004}, silicate absorption has been observed from 8--11\,$\umu$m in brown dwarfs of spectral type L1--L8 \citep{Roellig2004, Cushing2006, Burgasser2008, Looper2008, Suarez2022, Suarez2023}. \citet{Suarez2022} show that this spectral type range spans the formation, thickening, and sedimentation of silicate condensate clouds in the atmospheres of ultracool dwarfs. The \textit{Spitzer} IRS spectra have also been used to determine the physical properties of the silicates, such as grain size and mineral composition, that can produce the observed features \citep[e.g.,][]{Burningham2021, Vos2023}. \citet{Luna2021} examined the predicted effects of these physical characteristics on the shape of silicate absorption features in their models of mid-IR spectra of brown dwarfs. These predictions have been used by \citet{Suarez2023} to further correlate variations in silicate absorption profiles among mid-L dwarfs as a function of atmospheric surface gravity and hence brown dwarf age.

Mid-IR \textit{JWST} observations have also been made for young planetary-mass objects, which appear spectroscopically similar to low-gravity brown dwarfs \citep[e.g.,][]{Miles2023, Patapis2025, Hoch2025}. Hence, spectral libraries and templates developed from observations of brown dwarfs, such as the aforementioned \textit{Spitzer} IRS spectra which form the basis of the ultracool dwarf spectral libraries from \citet{Cushing2006} and \citet{Suarez2022}, can have significant utility when analysing the spectra of young planetary-mass objects.

While there are a few brown dwarfs with relatively high S/N ratios in the \textit{Spitzer} IRS data set, most range between S/N of \hbox{10--50}. The Low Resolution Spectroscopy (LRS) mode on the \textit{JWST} mid-IR Instrument (MIRI) offers an opportunity for improvement, with sensitivities $\sim 50\times$ greater than those of \textit{Spitzer} IRS over a similar wavelength range and at similar spectral resolution of $R \equiv \Delta \lambda /\lambda \sim 100$ \citep{Rieke2015}.

Here, we present a new mid-IR spectral library of brown dwarfs observed with MIRI LRS. We discuss our sample selection, data reduction, and identification of new bona fide or candidate binaries in $\S$\ref{sec:sampleanddata}, followed by a presentation and discussion of our spectral library in $\S$\ref{sec:spectra}. We then explore the spectroscopic diversity of our observed sample and discuss individual dwarfs with peculiar spectral features in $\S$\ref{sec:withintypecomparison}. We examine trends in the absolute magnitudes of the dwarfs in our observed sample in $\S$\ref{sec:absolutemagnituderelations} before summarizing our findings in $\S$\ref{sec:conclusion}.

\section{Sample selection and observations}
\label{sec:sampleanddata}

\subsection{Parent and observed target samples}
\label{sec:sampleselection}

GO 3930 (PI: S.\ Metchev) was a \textit{JWST} MIRI LRS survey program with a parent sample of 132 M9--T6.5 dwarfs, of which a subset of 23 L0--T6 dwarfs were successfully observed.
Survey programs include a large list of potential targets, with a subset of this list observed to fill gaps in the \textit{JWST} observing schedule. The majority (106 targets) of the parent sample was selected from compilations of sources with well-characterised photometric variability, projected rotational velocities ($v\sin i$), and trigonometric parallax \citep{Buenzli2014, Metchev2015, MilesPaez2017, MilesPaez2023, Tannock2023}. The remaining 26 targets in the parent sample were selected from the UltracoolSheet (UCS) compendium of $>$4000 ultracool dwarfs \citep{Best2021, Best2025}, and were chosen to improve coverage at high ecliptic latitudes; a subset of these targets are similarly well-characterised to those above. Given the best-practice recommendations for \textit{JWST} survey-type programs, a full-sky target distribution was desirable to maximize opportunities for observation, while high ecliptic latitudes alleviated \textit{JWST} Micrometeoroid Avoidance zone (MAZ\footnote{\url{https://jwst-docs.stsci.edu/jwst-observatory-characteristics-and-performance/jwst-target-viewing-constraints/jwst-micrometeoroid-avoidance-zone}}) constraints. 

Where high-angular resolution data were available from adaptive optics-assisted or \textit{Hubble Space Telescope (HST)} observations, the targets were vetted against resolved binarity. This was possible for about 50 per cent of the parent sample of 132 targets, including 10 of the 23 sources that were observed. Confirmed members of known young stellar moving groups were also excluded, based on their likely stellar association listed in the UCS. However, targets that were overluminous in the near- or mid-IR
with no other evidence of binarity, or that were not known members of young stellar moving groups, or that had otherwise peculiar spectroscopic features were not removed, in order to avoid discrimination against the diversity of ultracool atmospheres. 

\begin{figure}
  \centering
  \includegraphics[width=\columnwidth, alt={Scatter plot showing colour versus absolute magnitude of brown dwarfs in this sample, showing overluminosity of candidate and resolved binaries as well as trends in colour for L and T dwarfs.}]{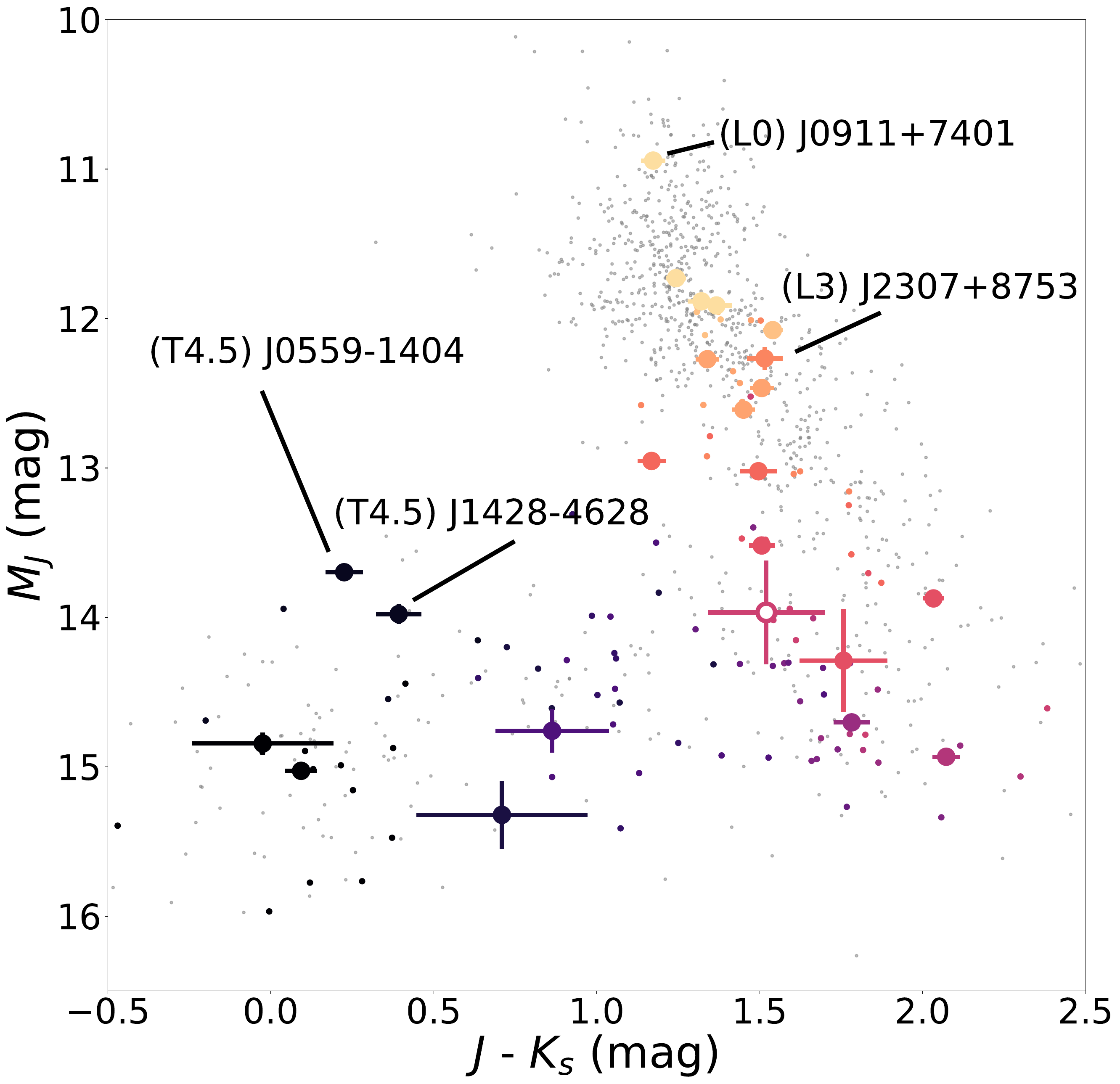}
  \caption{$M_J$ vs. $J-K_s$ colour-magnitude diagram of the 23 dwarfs observed in our \textit{JWST} survey program (large coloured points), the 123 dwarfs in our parent sample (small colour points), and the full UCS sample of dwarfs with trigonometric parallaxes (small grey points). Colouring is based on spectral type. Our observed sample contains one dwarf without a trigonometric parallax (J1731+5310, L6), which is shown with a hollow marker. Candidate or newly resolved binaries as discussed in Section \ref{sec:youngandbinary} are labelled.}
  \label{fig:sampleCMD}
\end{figure}

The parent sample of 132 targets was spread nearly uniformly across all spectral types between M9--T6, with the intention of complementing existing \textit{JWST} observations of $\geq$T6 dwarfs \citep{Beiler2024}. The 23 observed targets have spectral types ranging from L0--T6, with $>$1 spectral sub-type gaps only between \hbox{L6--L7.5}, L8--T1.5, and T1.5--T3.5 (Table \ref{table:observations}). We show the observed sample as well as the full parent sample on a near-infrared (NIR)
2MASS $M_J$ vs.\ $J-K_s$ colour-magnitude diagram in Fig. \ref{fig:sampleCMD}. When discussing the dwarfs in our sample, we use shortened versions of their J2000 identifiers listed in Table \ref{table:observations} (e.g., 2MASS J05591914$-$1404488 as J0559$-$1404).

\subsubsection{Candidate binary and young dwarfs}
\label{sec:youngandbinary}

As we do not remove overluminous dwarfs, our observed sample of 23 targets includes four overluminous candidate binaries. The T4.5 dwarfs J1428$-$4628 and J0559$-$1404 each have prior published indications of significant NIR overluminosity, 
suggestive of unresolved binarity \citep[][respectively]{Lodieu2014, Burgasser2001}. The L0 dwarf J0911+7401 and L3 dwarf J2307+8753 are also overluminous in the NIR but do not have previous published discussion of this overluminosity. We resolve J2307+8753 and J1428$-$4628 into binaries for the first time in our target acquisition images (Section \ref{sec:resolvedbinaries}). As these are both marginally resolved systems, we are unable to separate the traces of the two components for independent analysis. The other two overluminous dwarfs remain unresolved and are discussed further in Section~\ref{sec:unresolvedbinaries}. We flag these four dwarfs as resolved or candidate binaries throughout our analysis.

\begin{landscape}
    \begin{table}
    \begin{center}
    \caption{L0--T6 dwarfs observed by the MIRI LRS spectral library GO 3930 program.}
    \label{table:observations}
    \begin{tabular}{c c c c c c c c c c c c c}
         \hline
         \hline
         Name & Short name & Spectral type & Spectral & W3 & Total int. & Observing & Observing start & Groups & Ints. per & Target acq. & Plx. & Plx. \\
         & & (Opt/NIR) & type ref(s) & (mag) & time (s) & date (UTC) & time (UTC) & per int. & exp. & filter & (mas) & ref(s) \\
         \hline
         SDSSp J053951.99$-$005902.0 & J0539$-$0059 & L5/L5 & (1)/(2) & 11.716 $\pm$ 0.288 & 1410 & 2024 Mar 17 & 05:35:16 & 50 & 5 & F560W & 78.9 $\pm$ 0.4 & (22) \\
         2MASS J05591914$-$1404488 & J0559$-$1404 & T5/T4.5 & (3)/(4) & 10.605 $\pm$ 0.089 & 1504 & 2024 Mar 25 & 11:36:04 & 15 & 17 & F1000W & 95.3 $\pm$ 0.7 & (22) \\
         2MASS J06050196$-$2342270 & J0605$-$2342 & L0/M9.8 & (5)/(6) & 12.648 $\pm$ 0.492 & 1393 & 2024 Mar 25 & 10:38:24 & 83 & 3 & F560W & 30.19 $\pm$ 0.27 & (22) \\
         2MASS J06395596$-$7418446 & J0639$-$7418 & L5/- & (5)/- & 12.519 $\pm$ 0.301 & 1393 & 2024 Mar 16 & 10:38:12 & 83 & 3 & F560W & 51 $\pm$ 8 & (23) \\
         DENIS-P J0652197$-$253450 & J0652$-$2534 & L0/L2$\beta$ & (7)/(8) & 10.366 $\pm$ 0.059 & 1521 & 2024 Apr 08 & 23:00:27 & 10 & 25 & F1000W & 62.26 $\pm$ 0.09 & (22) \\
         WISEA J071552.38$-$114532.9 & J0715$-$1145 & -/L4pec & -/(9) & 11.756 $\pm$ 0.242 & 1410 & 2024 Mar 17 & 12:14:33 & 50 & 5 & F560W & 55.5 $\pm$ 0.3 & (22) \\
         2MASS J07193188$-$5051410 & J0719$-$5051 & L0/L0 & (10)/(11) & 11.698 $\pm$ 0.156 & 1410 & 2024 Mar 11 & 23:04:23 & 50 & 5 & F560W & 36.17 $\pm$ 0.19 & (22) \\
         2MASSI J0755480+221218 & J0755+2212 & T6/T5 & (3)/(4) & $>$11.761 & 1410 & 2024 Mar 28 & 15:52:44 & 50 & 5 & F560W & 66.57 $\pm$ 1.20 & (24) \\
         DENIS J081730.0$-$615520 & J0817$-$6155 & -/T6 & -/(12) & 9.661 $\pm$ 0.033 & 1548 & 2024 Mar 12 & 01:54:24 & 7 & 35 & F1500W & 191.8 $\pm$ 0.4 & (22) \\
         2MASSI J0825196+211552 & J0825+2115 & L7.5/L7pec & (13)/(2) & 10.509 $\pm$ 0.096 & 1504 & 2024 Mar 28 & 17:53:52 & 15 & 17 & F1000W & 92.6 $\pm$ 0.8 & (22) \\
         SSSPM J0829$-$1309 & J0829$-$1309 & L2/L2 & (14)/(11) & 10.092 $\pm$ 0.052 & 1521 & 2024 Apr 10 & 14:05:03 & 10 & 25 & F1000W & 85.6 $\pm$ 0.14 & (22) \\
         2MASSI J0835425$-$081923 & J0835$-$0819 & L5/L4pec & (15)/(16) & 9.472 $\pm$ 0.039 & 1659 & 2024 May 04 & 14:32:32 & 5 & 50 & F1500W & 138.31 $\pm$ 0.21 & (22) \\
         SDSS J090900.73+652527.2 & J0909+6525 & -/T1.5 & -/(17) & 11.044 $\pm$ 0.197 & 1437 & 2024 Mar 03 & 08:29:11 & 25 & 10 & F560W & 55.60 $\pm$ 2.99 & (24) \\
         2MASS J09111297+7401081	& J0911+7401 & L0/- & (10)/- & 10.559 $\pm$ 0.076 & 1504 & 2024 Mar 05 & 07:28:04 & 15 & 17 & F1000W & 40.23 $\pm$ 0.13 & (22) \\
         2MASS J10511900+5613086	& J1051+5613 & L2/L0.8 & (10)/(6) & 10.782 $\pm$ 0.095 & 1504 & 2024 Mar 17 & 21:15:31 & 15 & 17 & F1000W & 63.92 $\pm$ 0.13 & (22) \\
         2MASSW J1108307+683017 & J1108+6830 & L1$\gamma$/L1.8 & (16)/(18) & 10.168 $\pm$ 0.049 & 1521 & 2024 May 06 & 14:22:36 & 10 & 25 & F1000W & 61.81 $\pm$ 0.13 & (22) \\
         2MASS J11263991$-$5003550 & J1126$-$5003 & L4.5/L6.5pec & (19)/(19) & 11.584 $\pm$ 0.146 & 1410 & 2024 Feb 08 & 15:18:57 & 50 & 5 & F560W & 61.83 $\pm$ 0.25 & (22) \\
         SDSS J121440.95+631643.4 & J1214+6316 & -/T3.5 & -/(17) & 12.254 $\pm$ 0.265 & 1415 & 2024 Mar 15 & 01:33:18 & 63 & 4 & F560W & 56 $\pm$ 5 & (25) \\
         2MASSW J1338261+414034 & J1338+4140 & L2.5/L2.4 & (13)/(6) & 11.711 $\pm$ 0.191 & 1410 & 2024 May 10 & 11:52:16 & 50 & 5 & F560W & 47.63 $\pm$ 0.22 & (22) \\
         WISE J140533.32+835030.5 & J1405+8350 & L8/L9 & (20)/(20) & 10.520 $\pm$ 0.058 & 1504 & 2024 Apr 08 & 04:28:42 & 15 & 17 & F1000W & 103.4 $\pm$ 0.5 & (22) \\
         2MASS J14284235$-$4628393$^\dagger$ & J1428$-$4628 & -/T4.5 & -/(21) & 12.022 $\pm$ 0.268 & 1415 & 2024 Mar 26 & 12:53:19 & 63 & 4 & F560W & 41.462 $\pm$ 0.018 & (22) \\
         SDSS J173101.41+531047.9 & J1731+5310 & -/L6 & -/(17) & 12.868 $\pm$ 0.355 & 1393 & 2024 Mar 28 & 00:11:28 & 83 & 3 & F560W & 31 $\pm$ 5* & (26) \\
         2MASS J23072655+8753294 & J2307+8753 & L3/- & (10)/- & 11.403 $\pm$ 0.118 & 1437 & 2024 Mar 26 & 01:32:35 & 25 & 10 & F560W & 32.8 $\pm$ 1 & (22) \\
         \hline
    \end{tabular}
    \end{center}
    $^\dagger$Alternatively, HIP 70849B.\\
    *Photometric parallax from \textit{ALLWISE} W2 photometry and spectral type vs.\ absolute magnitude relation in \citet{Dupuy2012}. \\
    References: (1) \citet{Fan2000}; (2) \citet{Knapp2004}; (3) \citet{Burgasser2003}; (4) \citet{Burgasser2006}; (5) \citet{Cruz2007}; (6) \citet{Bardalez2014}; (7) \citet{Phan2008}; (8) \citet{Bardalez2019}; (9) \citet{Kirkpatrick2014}; (10) \citet{Reid2008}; (11) \citet{Marocco2013}; (12) \citet{Artigau2010}; (13) \citet{Kirkpatrick2000}; (14) \citet{Lodieu2005}; (15) \citet{Cruz2003}; (16) \citet{Gagne2015}; (17) \citet{Chiu2006}; (18) \citet{Aller2016}; (19) \citet{Burgasser2008}; (20); \citet{Castro2013}; (21) \citet{Lodieu2014}; (22) \citet{Gaia2023}; (23) \citet{Smart2018}; (24) \citet{Vrba2026}; (25) \citet{Kirkpatrick2021}; (26)
    \citet{Best2025}
    \end{table}
\end{landscape}

Two more targets have evidence of binarity discussed in the literature: the T6 dwarf J0817$-$6155 with marginal astrometric evidence of binarity \citep{Cheng2025} and the T3.5 dwarf J1214+6316 as a possible composite NIR spectral binary \citep{Geissler2011}. We discuss these two T dwarfs further in Section~\ref{sec:T1-T6comparison}, and consider them only as possible binaries. We list our newly resolved, candidate, and possible binaries in Table \ref{table:binaries}.

\begin{table*}
    \caption{Newly resolved, candidate, and possible binaries based on our analysis.}
    \begin{tabular}[width=\columnwidth]{|c c c c c|}
        \hline
        \hline
        Name & Reason for suspected binarity & Flux ratio (per cent) & Angular separation (arcsec) & Position angle (deg) \\
        \hline
        \multicolumn{5}{|c|}{\textbf{Resolved}} \\
        \hline
        2MASS J14284235$-$4628393 (T4.5) & Resolved in F560W target & 58 $\pm$ 4 & 0.124 $\pm$ 0.006 & 283 $\pm$ 2 \\ & acquisition imaging \\
        2MASS J23072655+8753294 (L3) & Resolved in F560W target & 76.1 $\pm$ 0.7 & 0.230 $\pm$ 0.002 & 341.3 $\pm$ 0.2 \\ & acquisition imaging \\
        \hline
        \multicolumn{5}{|c|}{\textbf{Candidate}} \\
        \hline
        2MASS J05591914$-$1404488 (T4.5) & Spectrophotometrically similar to & ? & < 0.05$^1$ \\ & newly resolved T4.5 binary \\ & 2MASS J14284235$-$4628393 \\
        2MASS J09111297+7401081 (L0) & Overluminous by factor of & ? & < 0.025$^2$ \\ & 2 compared to dwarfs of\\ & same spectral type \\
        \hline
        \multicolumn{5}{|c|}{\textbf{Possible}} \\
        \hline
        DENIS J081730.0$-$615520 (T6) & Moderately overluminous, & ? & < 0.25$^3$ \\ & elevated \textit{Gaia} RUWE \\
        SDSS J121440.95+631643.4 (T3.5) & Spectroscopically consistent & ? & < 0.1$^3$ \\ & with combination of T1.5\\ & and T6 dwarfs\\
        \hline
        
    \end{tabular}
    \label{table:binaries}
\begin{flushleft}
$^1$\citet{Burgasser2003}.\\
$^2$\citet{Factor2022}.\\
$^3$MIRI target acquisition imaging.
\end{flushleft}
\end{table*}

In addition, while none of the dwarfs in our observed sample is a known member of $\lesssim$100 Myr a young moving groups, there are published indications of youth for four of our dwarfs. The L0 dwarf J0652$-$2534 shows Li I absorption and optical low gravity features \citep{Burgasser2015}, and was given a $\beta$ NIR gravity classification by \citet{Bardalez2019}. The L5 dwarf J0835$-$0819 was kinematically assigned to the 625 Myr-old Hyades by \citet{Seifahrt2010}, and noted as peculiar (though not due to low gravity) by \citet{Gagne2015}. \citet{Liu2016} find NIR spectroscopic evidence of intermediate gravity for this dwarf. Another L0 dwarf, J0911+7401, shows evidence for youth based on \textit{GALEX} near-UV overluminosity and kinematics \citep{Lee2022}. The L1 dwarf J1108+6830 shows NIR and optical spectral signatures of low gravity and lies only marginally outside the Carina young moving group in both position and velocity space \citep{Gagne2015}. We flag these four dwarfs as potentially young throughout this paper.

\subsection{Observations}
\label{sec:observations}

Our targets were observed using the MIRI LRS mode of \textit{JWST}. We used a standard 2-point dither with the FASTR1 readout mode for spectroscopic observations. For target acquisition, we used the bluest of the F560W, F1000W, or F1500W filters that avoided saturation. Survey programs with \textit{JWST} are recommended to have total observing times at or under 90 minutes per target. To maximise the S/N and archival value, we used all available time remaining per observation once overheads were accounted for, while keeping each visit to 90 minutes total. 

Total integration times consequently ranged from 1390--1660 s per target. All targets are brighter than 12.5 mag in the \textit{WISE} W3 (11.6\,$\umu$m) band, enabling us to obtain observations with S/N $>$50 at the $\sim$12\,$\umu$m red end of the silicate feature. In Table \ref{table:observations}, we list the observed L0--T6 dwarfs, along with their spectral types, \textit{WISE} W3 magnitudes from \citet{Cutri2021}, and the properties of their observation (exposure time, observing date, groups per integration, integrations per exposure, and filter used in target acquisition imaging). Optical and NIR spectral types were taken from the UCS, with the corresponding references listed in Table \ref{table:observations}. Except where the respective spectral type was unavailable, optical spectral type was used for L dwarfs and NIR spectral type for T dwarfs, in keeping with the classification schemes described in, e.g., \citet{Burgasser2002}. 

We also include parallax data for each target in Table \ref{table:observations}. 18 out of 23 sources in our observed sample have \textit{Gaia} parallax values, while an additional 4 sources have parallaxes from \textit{Spitzer} \citep{Kirkpatrick2021}, ESO \citep{Smart2018}, or USNO \citep{Vrba2026}. The remaining target, L6 dwarf J1731+5310, does not have published parallax data, so its nominal distance of 33 $\pm$ 5 pc is calculated directly from its \textit{ALLWISE} W2 photometry using the spectral type-absolute magnitude relation from \citet{Dupuy2012}. Obtaining high-quality parallax measurements for this dwarf should be a priority to maximize its utility as a spectral standard.

In addition to the observations listed in Table \ref{table:observations}, we attempted observations of the brown dwarfs 2MASS~J06453153$-$6646120 (sdL8), WISEA~J082640.45$-$164031.8 (L9), 2MASSI~J0859254$-$194926 (L7), and 2MASS~J12373919+6526148 (T6.5). However, target acquisition for each of these dwarfs was unsuccessful, with each target falling outside the region on the detector where \textit{JWST}'s centroiding algorithm can correct for small errors in provided coordinates. This was likely a result of these dwarfs' relatively high and uncertain proper motions. The targets fall outside the detector imaging region in target acquisition images, and the 2D spectra on the LRS region of the detector do not contain a dispersed source. We used these high-fidelity blank sky spectra for an improved bad pixel mask correction in Section \ref{sec:badpixels}.

\subsection{Data reduction}
\label{sec:datareduction}
The observations were processed using version 2.0.0 of the \textit{JWST} data reduction pipeline and reference file context jwst$\_$1535.pmap. We used the default (box) extraction with local background subtraction performed using the mean background fit across two five-pixel-wide windows on either side of the spectral trace. Default parameter values were used for all other pipeline stages.

Version 1.18.0 of the pipeline introduced the option of optimal extraction for LRS observations. With the default PSF reference file, we found that the optimal extraction mostly reproduced the spectra from the default box extraction. However, there is a persistent issue where the reference PSF is not accurately centred at long wavelengths, resulting in a systematic undercounting of flux at the long-wavelength end. Since the optimal extraction is still a work in progress, we chose to proceed with the more well-tested default box extraction. However, we note that the agreement with the optimal extraction throughout the nominal wavelength range ($\lesssim$12\,$\umu$m) reinforces the conclusions we draw from analysing these spectra.

\subsubsection{Additional bad pixel masking}
\label{sec:badpixels}
Between stages 1 and 2 of the pipeline, we introduced an additional step of supplementary bad pixel masking. The default \textit{JWST} data reduction pipeline contains functionality to handle bad pixels, but the existing maps of defective pixels used as reference files are incomplete. A small number of unflagged bad pixels impinge directly on the spectral extraction region of the detector, resulting in artefacts in the final products.

To remedy this, we made use of the deep blank-sky LRS observations obtained from the four aforementioned targets with failed target acquisition, as discussed in Section~\ref{sec:observations}. By combining these observations, we were able to isolate persistent unmarked bad pixels on the LRS detector. We then used a sliding window to isolate outlier pixels at a 3$-\sigma$ level, which were added to a mask that marks these pixels for handling by the existing pipeline in stage 2 of the data reduction.

\begin{figure}
  \centering
  \includegraphics[width=\columnwidth, alt={Graph showing spectra processed with and without the new bad pixel mask, including an inset highlighting the most pronounced improvement.}]{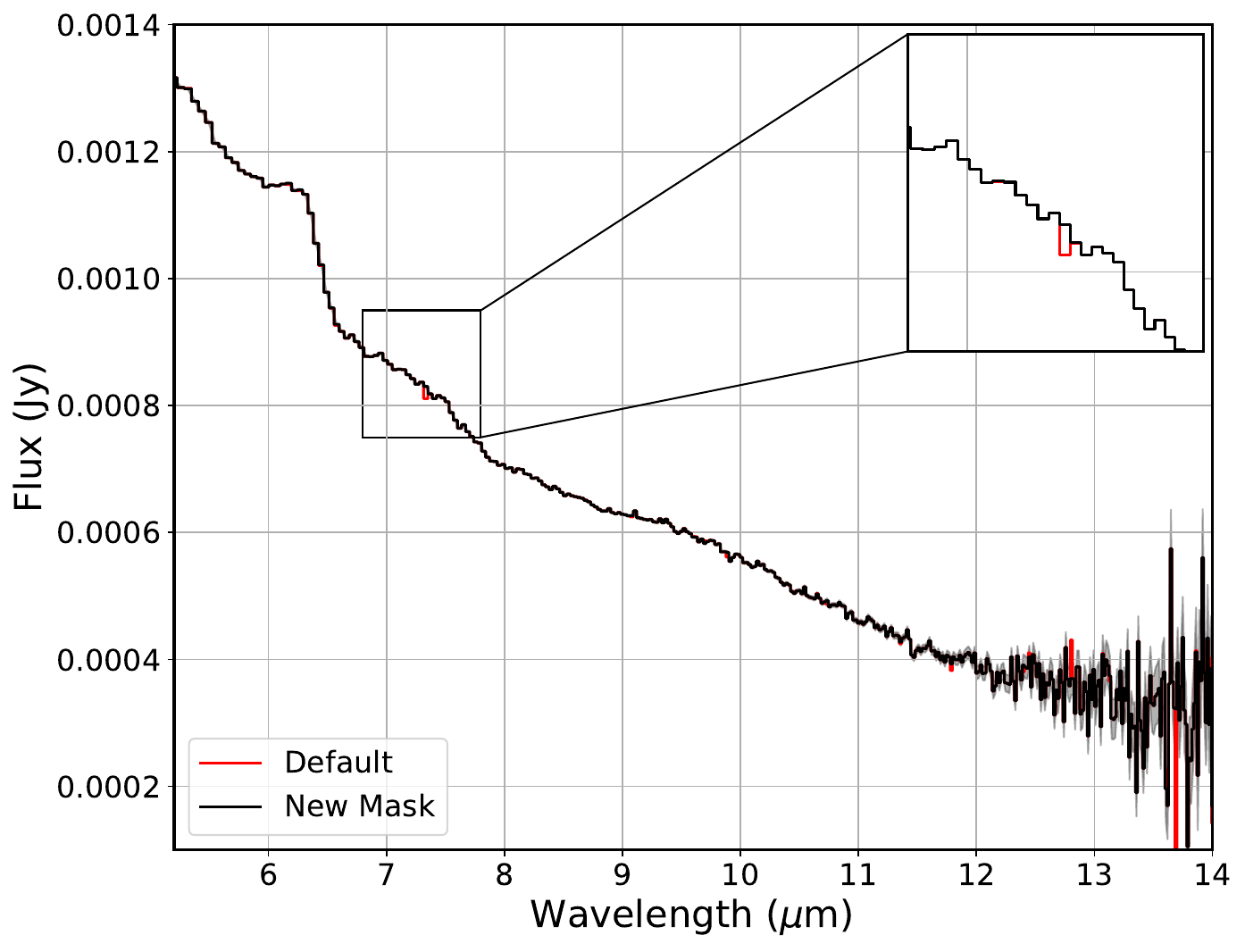}
  \caption{Comparison of spectra of the L0 dwarf J0605$-$2342 that use the default bad pixel mask (red) and our custom improved bad pixel mask (black; Section~\ref{sec:badpixels}). We note minor improvements near 7.3\,$\umu$m, 12.8\,$\umu$m, and 13.7\,$\umu$m.}
  \label{fig:badpixelcomparison}
\end{figure}

Fig.\ \ref{fig:badpixelcomparison} shows the improvement in the extracted spectra when the additional bad pixel masking is applied. These improvements are observed consistently across our spectral library. The most prominent effects are observed near 7.3\,$\umu$m, 12.8\,$\umu$m, and 13.7\,$\umu$m. Each of these differences between the default mask and our new mask corresponds to bad pixels that impinge directly on the spectra. 

\subsubsection{Supplemental MIRI photometry}
\label{sec:photometry}

For each target in our observed sample, a MIRI target acquisition image was taken in one of three filters (F560W, F1000W, F1500W; see Table~\ref{table:observations}) and processed through the \textit{JWST} data reduction pipeline. These images are flux-calibrated, so can be used for photometric purposes. We use the Python-based photometry package \textsc{Photutils} \citep{Bradley2025} and the simulated \textit{JWST} point-source function (PSF) contained within STPSF (formerly WebbPSF, \citealt{Perrin2012}) to perform PSF photometry on our target acquisition images.

We first fit a single PSF to each source using the PSF fitting functions found in the \textsc{Photutils} package. The best-fitting PSF is then used to extract calibrated flux in a given filter, which we add to our overall spectrophotometry. As the existing STPSF reference PSFs do not account for detector effects such as the MIRI brighter-fatter effect \citep{Argyriou2023}, the fit for each source is imperfect and leaves residuals of up to 1.5 per cent the source flux. We find that these residuals are well-contained within a circular region with radius 2$\times$ the FWHM of the respective imaging band pass. As such, we extract the residual flux using a circular aperture of this size, and add the result to our PSF-fitted flux. Due to the small contribution of these residuals, the corresponding increase in uncertainty is negligible. 

We compute the fit quality $q_{fit}$, defined as the ratio of the residual flux within the PSF region to the extracted flux, with the \textsc{Photutils} PSF photometry tools. For previous \textit{JWST} and \textit{HST} photometric analyses, a good PSF fit has been defined as $q_{fit} < 0.05$ \citep{Libralato2024}. The majority of the dwarfs in our observed sample are adequately fit according to this metric by a single-component PSF. We find we are able to probe separations down to $\sim0.5 \times$ filter FWHM in our target acquisition imaging. These dwarfs are thus unresolved to separations of 0.104 arcsec for objects observed with F560W, 0.164 arcsec for F1000W, and 0.244 arcsec for F1500W \citep{Dicken2024}. 

Three of the four sources with failed target acquisitions were not recovered in the target acquisition imaging. All three have high proper motions, $\geq 950$ mas/yr. As such, uncertainties in their proper motions may have compounded to place them too far from their expected locations for the \textit{JWST} pointing system to compensate. The remaining source with a failed acquisition, J0859$-$1949, is recovered within 1 arcsec of its expected location and $\sim$2 arcsec away from the expected target acquisition region of the MIRI detector readout. It is unclear why target acquisition failed in this case, however since the source still falls on the MIRI detector we apply the same photometry techniques as for the remainder of our observed sample.

We list the photometry for each dwarf in our observed sample, including the filter used and any additional notes, in Table \ref{table:photometry}.

\begin{table}
    \caption{\textit{JWST} MIRI acquisition-image photometry of dwarfs in our sample.}
    \begin{tabular}[width=\columnwidth]{|c c c c|}
        \hline
        \hline
        Name & Filter & Apparent magnitude & Notes \\
        \hline
        J0539$-$0059 & F560W & 11.409 $\pm$ 0.004 & \\
        J0559$-$1404 & F1000W & 10.546 $\pm$ 0.003 & \\
        J0605$-$2342 & F560W & 12.482 $\pm$ 0.006 & \\
        J0639$-$7418 & F560W & 12.770 $\pm$ 0.006 & \\
        J0652$-$2534 & F1000W & 10.292 $\pm$ 0.002 & \\
        J0715$-$1145 & F560W & 11.964 $\pm$ 0.004 & \\
        J0719$-$5051 & F560W & 12.115 $\pm$ 0.005 & \\
        J0755+2212 & F560W & 13.673 $\pm$ 0.009 & \\
        J0817$-$6155 & F1500W & 9.389 $\pm$ 0.003 & \\
        J0825+2115 & F1000W & 10.731 $\pm$ 0.003 & \\
        J0829$-$1309 & F1000W & 10.185 $\pm$ 0.002 & \\
        J0835$-$0819 & F1500W & 9.284 $\pm$ 0.003 & \\
        J0859$-$1949 & F560W & 12.112 $\pm$ 0.004 & Failed target \\ & & & acquisition \\
        J0909+6525 & F560W & 13.383 $\pm$ 0.008 & \\
        J0911+7401 & F1000W & 10.535 $\pm$ 0.003 & \\
        J1051+5613 & F1000W & 10.766 $\pm$ 0.003 & Possible cosmic ray \\ & & & contamination \\
        J1108+6830 & F1000W & 10.226 $\pm$ 0.002 & \\
        J1126$-$5003 & F560W & 11.840 $\pm$ 0.004 & \\
        J1214+6316 & F560W & 13.380 $\pm$ 0.008 & \\
        J1338+4140 & F560W & 11.909 $\pm$ 0.004 & \\
        J1405+8350 & F1000W & 10.584 $\pm$ 0.003 & \\
        J1428$-$4628 & F560W & 14.13 $\pm$ 0.04 (A) & System magnitude: \\
        & & 14.73 $\pm$ 0.06 (B) & 13.64 $\pm$ 0.07 \\
        J1731+5310 & F560W & 13.547 $\pm$ 0.009 & \\
        J2307+8753 & F560W & 12.767 $\pm$ 0.006 (A) & System magnitude: \\ & & 13.063 $\pm$ 0.006 (B) & 12.152 $\pm$ 0.009 \\
        \hline
        
    \end{tabular}
    \label{table:photometry}
\end{table}

\subsubsection{Newly resolved binaries}
\label{sec:resolvedbinaries}

\begin{figure}
  \centering
  \includegraphics[width=\columnwidth, alt={Imaging and point source function fitting residuals for two newly resolved binaries, showing visible residuals with one-source fitting and no clear residuals with two-source fitting.}]{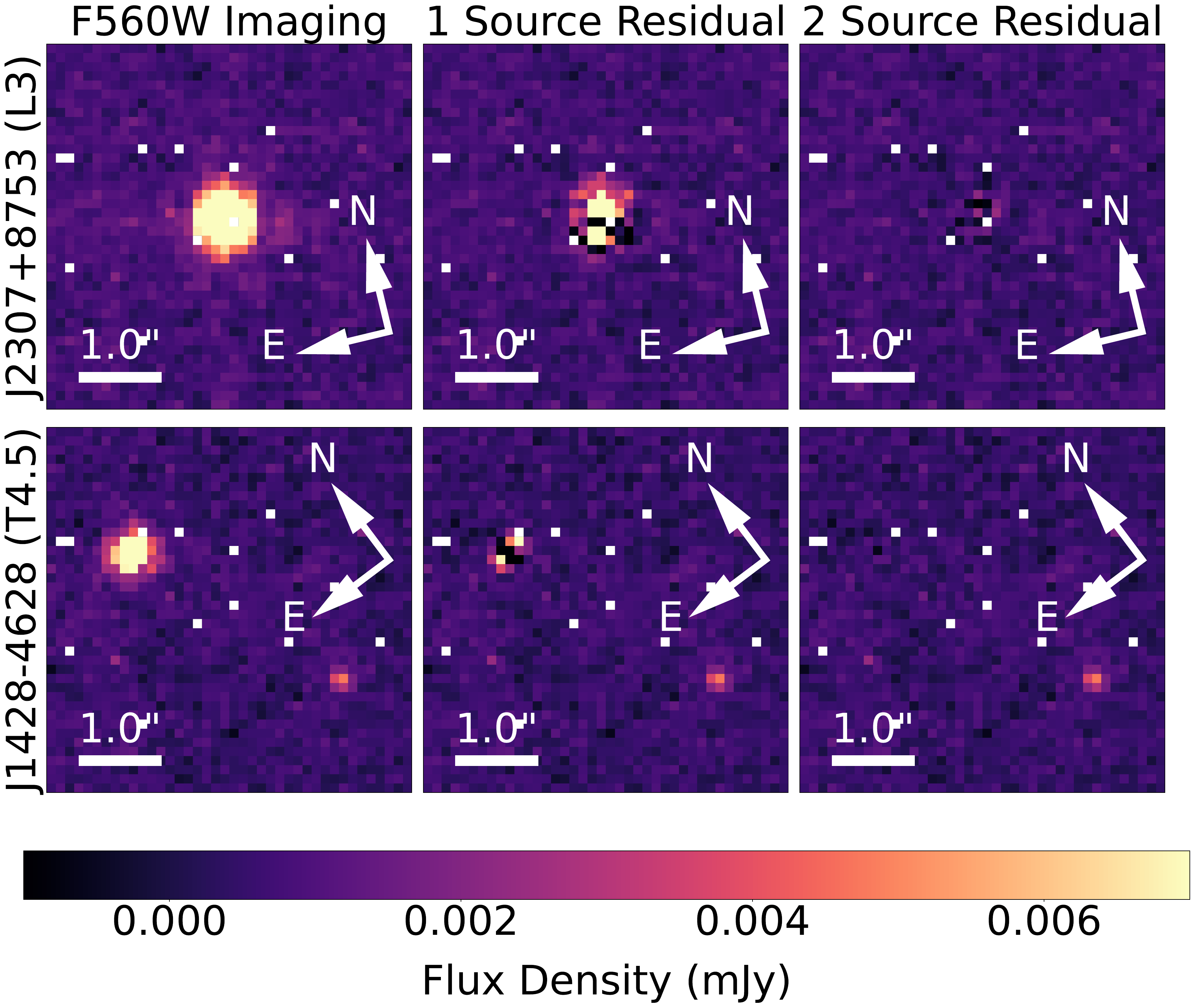}
  \caption{PSF fits to the L3 dwarf J2307+8753 and T4.5 dwarf J1428$-$4628. The left column shows the MIRI target acquisition imaging of the respective dwarfs. The middle column shows the residuals when the best single-source fit is subtracted, while the right column shows the residuals from the best two-source fit. The additional object in frame in the lower images is a background object, confirmed by the lack of shared proper motion in comparison of our target acquisition images to archival 2MASS imaging.}
  \label{fig:binaryresiduals}
\end{figure}

For two of our targets, J2307+8753 (L3) and J1428$-$4628 (T4.5), a single PSF fit leaves significant localized residuals, suggesting a better fit may be obtained with a two-component fit. \textsc{Photutils} allows the simultaneous fitting of multiple components, enabling us to iteratively fit a secondary component over a subpixel-sampled grid of separations in both axes. For these two sources, the fit quality is increased by a factor of 8.0 and 2.1, respectively, when a secondary component is added, and the visible residuals are eliminated. We show the single- and two-source fits to each source in Fig. \ref{fig:binaryresiduals}.

The best-fitting two-component solution to J2307+8753 (L3) gives components with apparent F560W magnitudes of 12.767 $\pm$ 0.006 mag and 13.063 $\pm$ 0.006 mag, corresponding to a secondary with 76.1 $\pm$ 0.7 per cent the flux of the primary at an angular separation of 0.230 $\pm$ 0.002 arcsec and a position angle of 341.3 $\pm$ 0.2 degrees. The best-fitting solution to J1428$-$4628 (T4.5) gives components with apparent F560W magnitudes of 14.13 $\pm$ 0.04 mag and 14.73 $\pm$ 0.06 mag, corresponding to a secondary with 58 $\pm$ 4 per cent the flux of its primary at an angular separation of 0.123 $\pm$ 0.006 arcsec and a position angle of 283 $\pm$ 2 degrees.

Neither of these dwarfs were previously known binaries, though J1428$-$4628 (T4.5) is itself a companion to the K7V star HIP 70849. There are some previous indications that brown dwarf + brown dwarf binaries are more common as wide companions to stellar hosts \citep{Burgasser2005}, with J1428$-$4628 extending this trend further. The parent star HIP 70849 is notable as an exoplanet host \citep{Segransan2011}, making HIP 70849 the second known system with this architecture of a brown dwarf pair as a wide companion to an exoplanet host star, after $\epsilon$ Indi \citep{McCaughrean2004, Feng2019, Subjak2023}. Additionally, in its discovery paper by \citet{Lodieu2014}, J1428$-$4628 was suggested to be a possible equal-mass binary, as it appears overluminous on an $M_J$ vs.\ $J-K_s$ NIR CMD (Fig. 1
of that work). However, no follow-up observations of this dwarf have been performed to attempt to resolve it to this point. J2307+8753 (L3) has no previous published indications of potential binarity. 

Due to the marginally resolved nature of these sources, we urge caution when working with their spectra. As mentioned in Section \ref{sec:youngandbinary}, this prevents separation of the traces of the two components for independent spectral analysis. Additionally, the \textit{JWST} data reduction pipeline applies corrections for slit losses, assuming a singular PSF at a known location in the slit; for resolved sources, this correction may be inadequate, particularly at longer wavelengths where an increased fraction of the PSF flux lies at greater distances from the nominal source location in the slit. J2307+8753 (L3), in particular is resolved along an axis nearly perpendicular to the slit, placing a larger fraction of the PSF for each component outside the slit compared to a binary resolved along the slit. This dwarf's spectrum shows a notably steeper slope compared to others of similar spectral type in our observed sample, as would be expected from slit loss correction deficiencies that become more pronounced at longer wavelengths as described above.

\subsubsection{Unresolved but overluminous candidate binaries}
\label{sec:unresolvedbinaries}

All other target acquisition images show no improvement in fit quality with two sources when compared to the single-source fit. This suggests that either these dwarfs are indeed single or they are unresolved at their respective imaging bands' limits. 

The other two candidate overluminous binaries, L0 dwarf J0911+7401 and T4.5 dwarf J0559$-$1404, described in Section \ref{sec:youngandbinary}, were each observed with filter F1000W for target acquisition, giving upper projected angular separation limits of 0.164 arcsec for an equal-flux component in each system. This is consistent with previous companion searches for each dwarf, which rule out near-equal luminosity companions down to 0.025 arcsec \citep{Factor2022} and 0.05 arcsec \citep{Burgasser2003} for J0911+7401 and J0559$-$1404, respectively. 
Additional constraints on the binarity of J0559$-$1404 exist from radial velocity monitoring, with which \citet{Prato2015} rule out companions in $\lesssim$1-day periods, corresponding to $\lesssim$0.003 au orbits, or $\lesssim$0.3 milliarcsec (mas) projected separations at the 10.5 pc distance to J0559$-$1404. These constraints leave an intermediate $\sim$0.3--50 mas angular separation range where a near-equal flux companion may still be present, which would require higher-angular resolution imaging or longer-duration RV monitoring to explore.

\section{Spectra and Features}
\label{sec:spectra}
\subsection{The MIRI LRS L0--T6 spectral library}
\label{sec:spectrallibrary}

\begin{figure*}
\includegraphics[width=0.85\paperwidth, alt={Graphs showing mid-infrared spectra of brown dwarfs, with absorption bands of water, silicates, ammonia, methane, and silicon monoxide overlaid in coloured blocks.}]{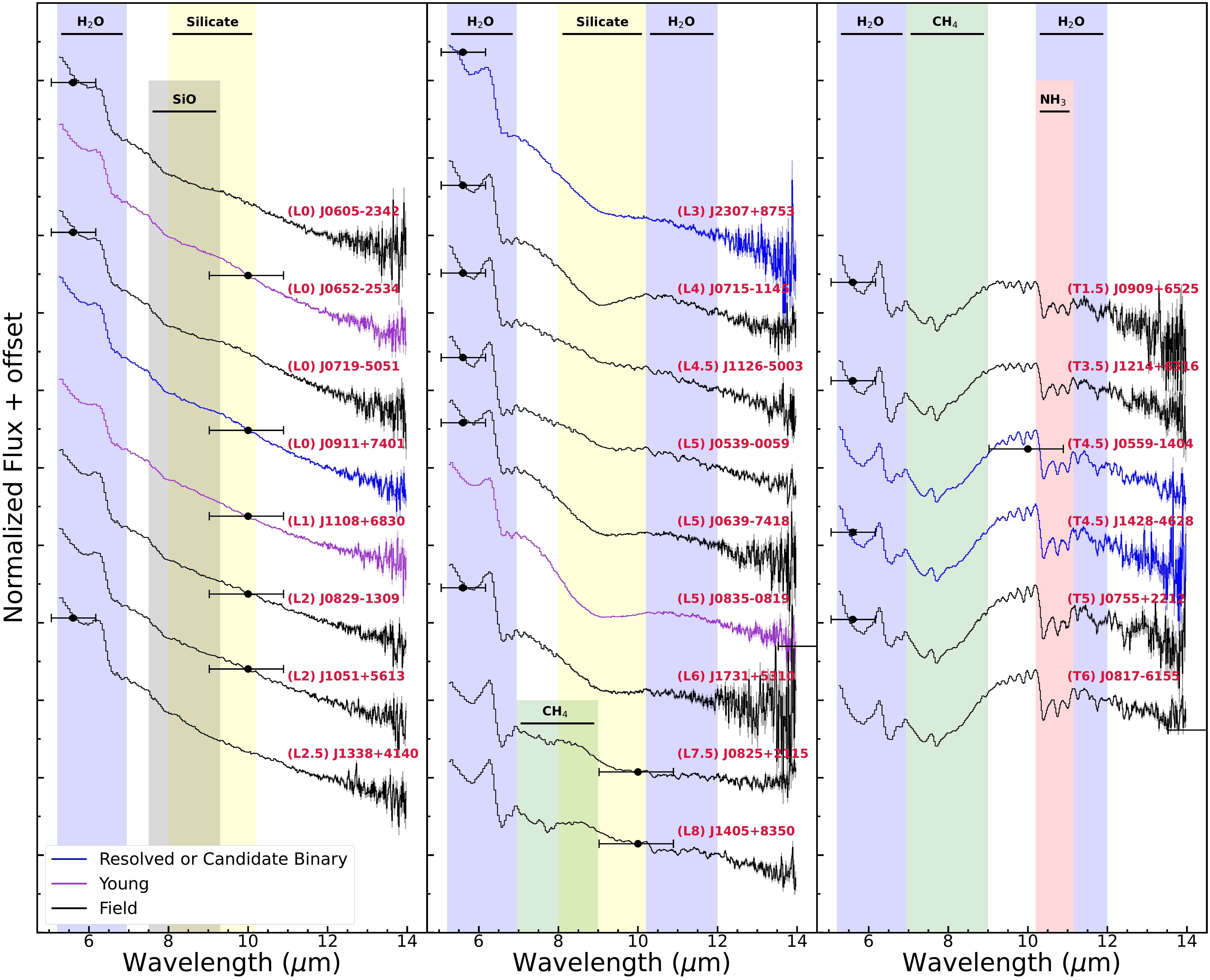}
\caption{Library of \textit{JWST} MIRI LRS 5.2--14.0 $\mu$m spectra of L0--T6 dwarfs. The left column contains L0--L2.5 dwarfs with minimal silicate absorption. The central column contains L3--L8 dwarfs with significant silicate presence in their spectra. The right column contains the T1.5--T6 dwarfs in the observed sample. Black data points show the MIRI acquisition image fluxes, with the extent of the horizontal errorbar corresponding to the width of the relevant filter (F560W, F1000W, or F1500W).
Four dwarfs in the observed sample are newly resolved or candidate binaries; the spectra for these dwarfs are shown in blue. For these, the black data points show the combined flux from the components, which matches the observed spectroscopic flux. Dwarfs with published indications of intermediate or low gravity are shown in purple. Coloured bands show regions of molecular or condensate absorption. The L0--L2.5 dwarfs consistently show SiO absorption, previously not identified in brown dwarfs.}
\label{fig:spectrallibrary}
\end{figure*}

Fig. \ref{fig:spectrallibrary} displays the main spectral library from our LRS observations, with all spectra normalized to the flux in the 11.7--12.3\,$\umu$m range and vertically shifted for clarity. The average SNR of the spectra in the library is $\sim$1000 at 5.2\,$\umu$m, decreasing to $\sim$90 at 12.0\,$\umu$m. The library includes eight L0--L2.5 dwarfs, nine L3--L8 dwarfs, and six T1.5--T6 dwarfs, shown in the left, middle, and right columns of Fig.~\ref{fig:spectrallibrary}. Four dwarfs in this sample are spatially resolved or candidate binaries (Section \ref{sec:youngandbinary}), and have their spectra shown in blue; caution should be taken when working with these spectra (see Section \ref{sec:resolvedbinaries} for more details).

\subsection{Molecular absorption indices: H$_2$O, NH$_3$, and CH$_4$}
\label{sec:absorptionindices}
We calculate absorption indices for various compounds using the Spectral Energy Distribution Analyzer (\textsc{SEDA}) package \citep{Suarez2021, Suarez2025}. This package includes modifications to the index formulas for gas molecules from \citet{Suarez2022}, originally developed by \citet{Cushing2006}, to better delineate between the flux regions for ammonia and methane. The indices for water, ammonia, and methane are calculated using the formulae from these works as, respectively,

\begin{equation}
\text{H} _2 \text{O Index} = \frac{F_{6.25}}{0.562 F_{5.80} + 0.474 F_{6.75}}
\label{eq:waterindex}
\end{equation}

\begin{equation}
\text{NH} _3 \text{ Index} = \frac{F_{9.9}}{F_{10.6}}
\label{eq:ammoniaindex}
\end{equation}

\begin{equation}
\text{CH} _4 \text{ Index} = \frac{F_{9.9}}{F_{7.65}}
\label{eq:methaneindex}
\end{equation}

\begin{figure}
  \centering
  \includegraphics[width=\columnwidth, alt={Graphs comparing water, ammonia, and methane absorption indices, calculated for Spitzer and James Webb mid-infrared spectra of brown dwarfs, showing general agreement between the data sets.}]{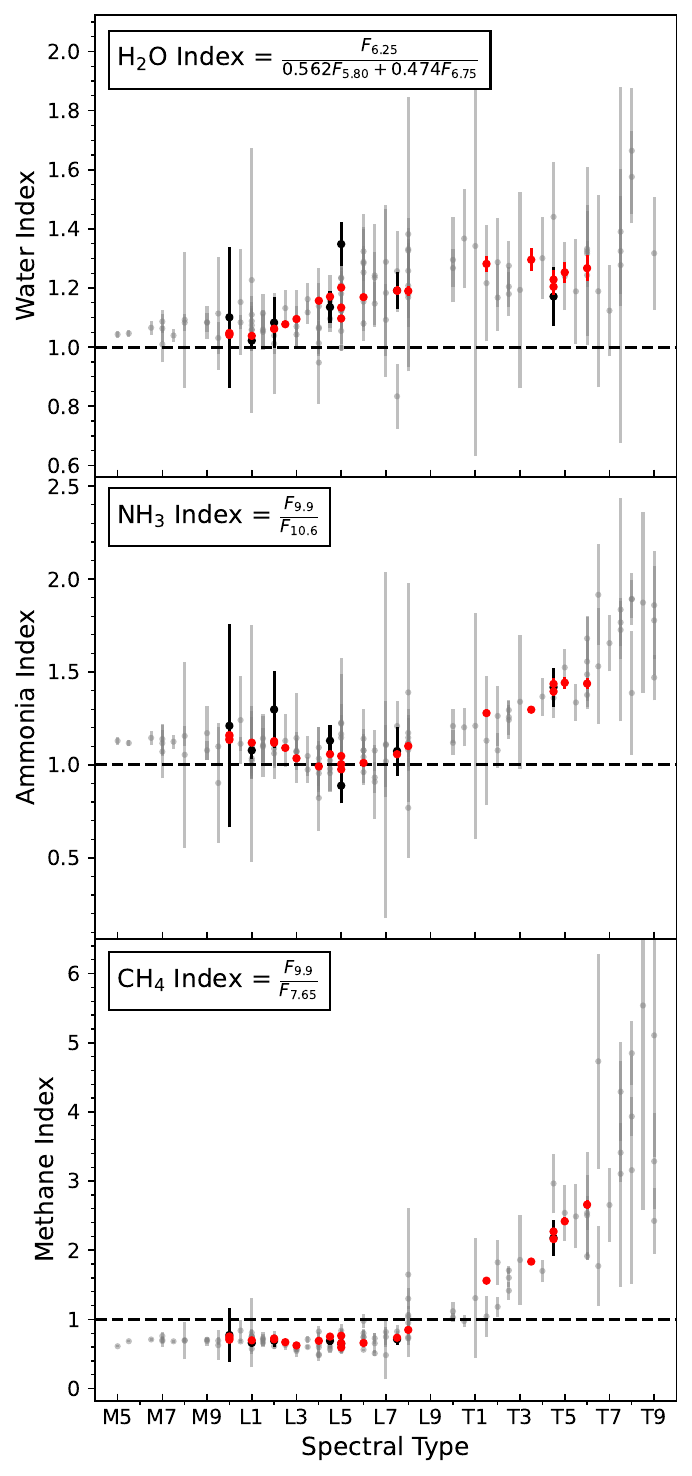}
  \caption{Water, ammonia, and methane indices for dwarfs in our MIRI LRS sample (red) as a function of spectral type. The grey points correspond to the spectral indices for the \textit{Spitzer} IRS spectra from \citet{Suarez2022}, calculated using Equations \ref{eq:waterindex}--\ref{eq:methaneindex} in this paper. \textit{Spitzer} indices for dwarfs also in our \textit{JWST} sample are shown in black.}
  \label{fig:molecularindices}
\end{figure}

$F_\lambda$ is defined as the mean flux density {over a certain wavelength range surrounding the indicated central wavelength, with a width of} 0.3\,$\umu$m for water or 0.6\,$\umu$m for methane and ammonia. Uncertainties are calculated based on the uncertainty of the mean flux for the absorption and continuum regions. We plot these indices as a function of spectral type in Fig. \ref{fig:molecularindices}. For comparison, the same indices are calculated for the \textit{Spitzer} IRS spectra from \citet{Suarez2022} and plotted on the same scale. Our observed sample contains seven dwarfs that are also a part of this \textit{Spitzer} IRS sample, including (L0) J0911+7401, (L1) J1108+6830, (L2) J1051+5613, (L4.5) J1126$-$5003, (L5) J0539$-$0059, (L7.5) J0825+2115, and (T4.5) J0559$-$1404. The \textit{Spitzer} IRS indices for these dwarfs are shown in black in Fig. \ref{fig:molecularindices}. 

We find that our MIRI LRS spectra produce index trends in agreement with those produced from \textit{Spitzer} IRS spectra, with a reduction in uncertainty such that most of our observed dwarfs have uncertainties in index value smaller than the data markers. Additionally, with the exception of the water index for J0539$-$0059 (L5), the \textit{Spitzer} and \textit{JWST} indices for the seven dwarfs in both samples agree within uncertainty.

\subsection{Silicate absorption: variations in depth, wavelength, and width}
\label{sec:silicateabsorption}

Silicate absorption is visible to varying degrees in all spectra from spectral type L0 (Section \ref{sec:L0sioandsilicate}; Fig. \ref{fig:L0comparison}) up to the L8 dwarf J1405+8350, marked with a yellow band in Fig. \ref{fig:spectrallibrary}. No obvious silicate absorption is visible in any of the T dwarf spectra, although without dedicated modelling we cannot rule out subtle absorption effects in the early-T dwarfs coinciding with the $\sim$10\,$\umu$m elevated flux region of the T dwarf spectra. This is in agreement with previous findings that place the dissipation of silicate clouds around the L-T transition \citep[e.g., ][]{Tsuji1999, Ackerman2001, Allard2001, Burgasser2002b, Knapp2004, Morley2012}. The trends in silicate absorption with spectral type are consistent with the observations of
\citet{Suarez2022, Suarez2023}
who find this absorption begins around spectral type L2 or L1 respectively, is strongest between L4--L6, and is absent beyond L8. 

The silicate absorption in all dwarfs in our observed sample from spectral types L2.5 to L6 is strongest at 9.0--9.1\,$\umu$m, consistent with the observations of field-age mid-L dwarfs in \citet{Suarez2023}. As the authors of that paper discuss, this central wavelength for the absorption is consistent with $\lesssim$0.1\,$\umu$m grains of enstatite or SiO dominating the silicate condensates. The L7.5 and L8 dwarfs show silicate absorption with a noticeably longer central wavelength than the other mid-L dwarfs, with maximum absorption at $\sim$9.4\,$\umu$m. As discussed by both \citet{Luna2021} and \citet{Suarez2023}, a longer maximum absorption wavelength can be produced by silicate grains of a larger size ($\sim$1\,$\umu$m) or more magnesium-rich composition compared to those observed in mid-L dwarfs. The modelling required to determine which specific species of silicate condensates produce our observed absorption features is beyond the scope of this paper, although such modelling has been performed in previous works \citep[e.g.,][]{Burningham2021, Vos2023, Molliere2025}. 

As with the molecular features, we use a silicate absorption index to quantify the depth of the silicate absorption in the dwarfs in our observed sample. We adopt the \citet{Suarez2022} definition of the silicate index, estimated from a linear fit to the fluxes in the 7.2--7.8\,$\umu$m and 11.2--11.8\,$\umu$m wavelength regions. The silicate index is then the ratio of the interpolated continuum flux at 9.0\,$\umu$m to the mean flux observed in a 0.6\,$\umu$m window around the same wavelength: 

\begin{equation}
\text{Silicate index} = \frac{C_{9.0}}{F_{9.0}}
\label{eq:silicateindex}
\end{equation}

We find that the silicate index as defined in \citet{Suarez2023} based on an absorption-free fitted exponential continuum underestimates the continuum flux at longer wavelengths, introduces additional scatter into the silicate indices, and has some difficulty fitting the shape of the continuum.

\begin{figure}
  \centering
  \includegraphics[width=\columnwidth, alt={Graph of a spectrum with visible silicate absorption, with shaded regions, a fitted trendline, and highlighted points showing the method of calculating the silicate index.}]{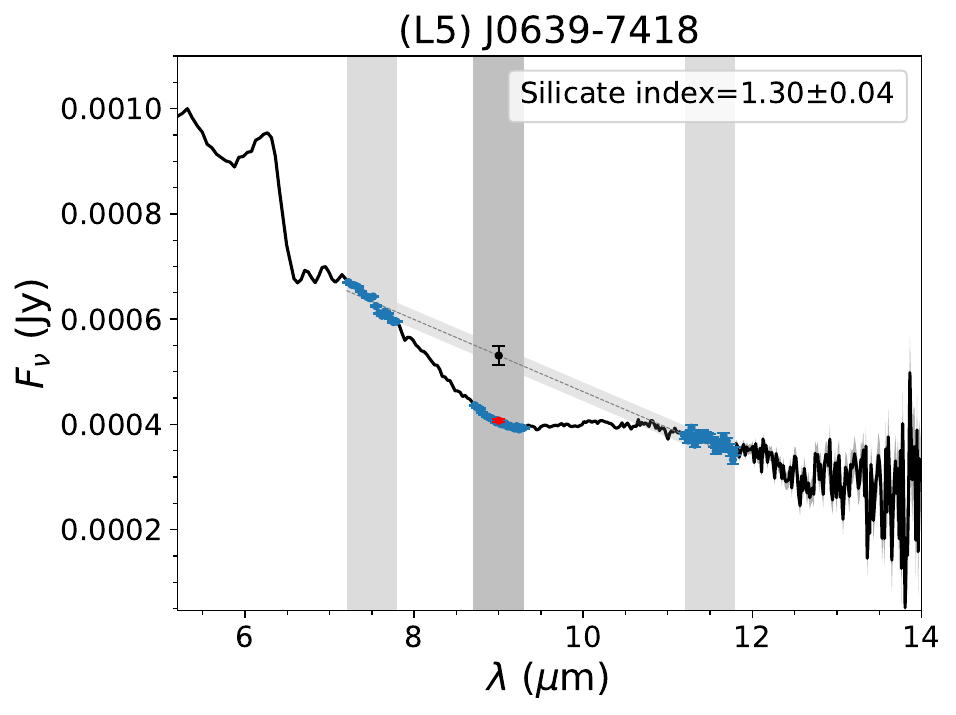}
  \caption{Example of silicate index calculation using the spectrum of the L5 dwarf J0639$-$7418, following \citet{Suarez2022}. The windows used to calculate the index are shaded in grey, with the darker region being the window for absorption and the lighter regions being the windows for continuum fitting. Series of blue data points with thicker error bars delineate the flux values used to calculate the continuum and the mean flux in the absorption region. The silicate index is defined as the ratio of the interpolated continuum flux (black point) to the absorption flux (red point).}
  \label{fig:silicateindexexample}
\end{figure}

\begin{figure}
  \centering
  \includegraphics[width=\columnwidth, alt={Graphs comparing silicate and silicon monoxide absorption indices, calculated for Spitzer and James Webb mid-infrared spectra of brown dwarfs, showing agreement between the data sets in the silicate index, while the Spitzer data are too noisy to observe silicon monoxide index trends.}]{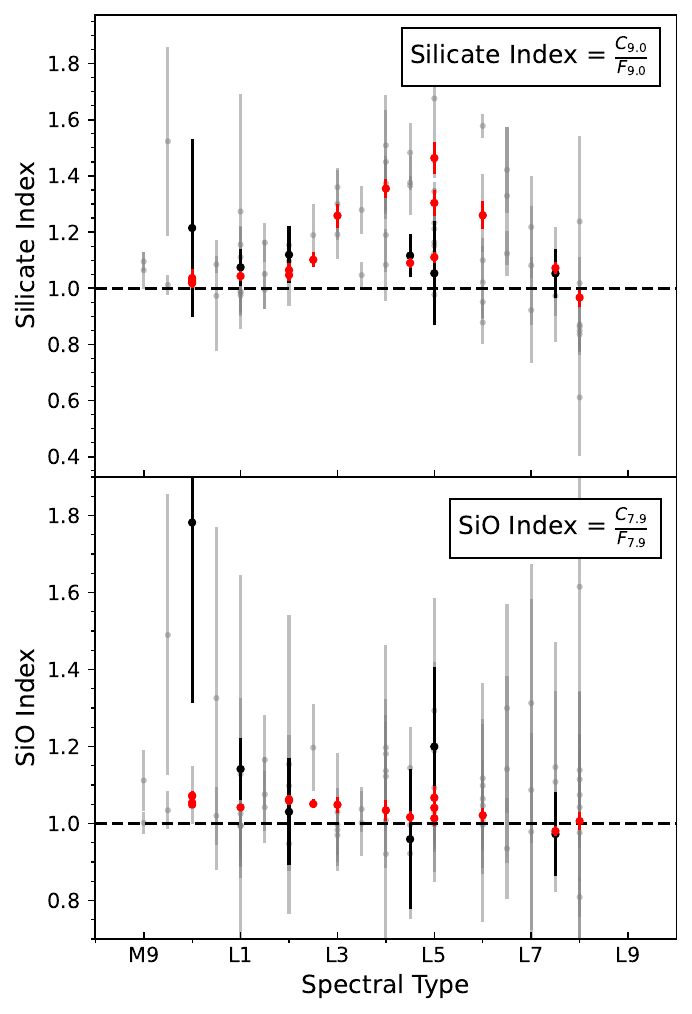}
  \caption{Silicate and SiO indices as defined in Sections \ref{sec:silicateabsorption} and \ref{sec:sioabsorption} respectively for dwarfs in our MIRI LRS sample (red) as a function of spectral type. The grey points correspond to the spectral indices for the \textit{Spitzer} IRS spectra from \citet{Suarez2022}, calculated using the Equations \ref{eq:silicateindex} and \ref{eq:sioindex} in this paper. \textit{Spitzer} indices for dwarfs also in our \textit{JWST} sample are shown in black.}
  \label{fig:silicateindices}
\end{figure}

We show an example of the silicate index calculation in Fig. \ref{fig:silicateindexexample}. The silicate indices for the dwarfs in our observed sample are plotted alongside those calculated similarly for the \textit{Spitzer} IRS data from \citet{Suarez2022} in the top panel of Fig. \ref{fig:silicateindices}. We only plot the results of our calculations for the L dwarfs, as the CH$_4$-induced modifications to the continuum that occur in T dwarfs make the indices meaningless for these objects. As with the molecular indices, we show the \textit{Spitzer} indices for dwarfs also in our \textit{JWST} sample in black.

\begin{figure}
  \centering
  \includegraphics[width=\columnwidth, alt={Graphs comparing silicate absorption profiles between mid-L brown dwarfs, including an inset scatter plot comparing quantitative characteristics of the absorption.}]{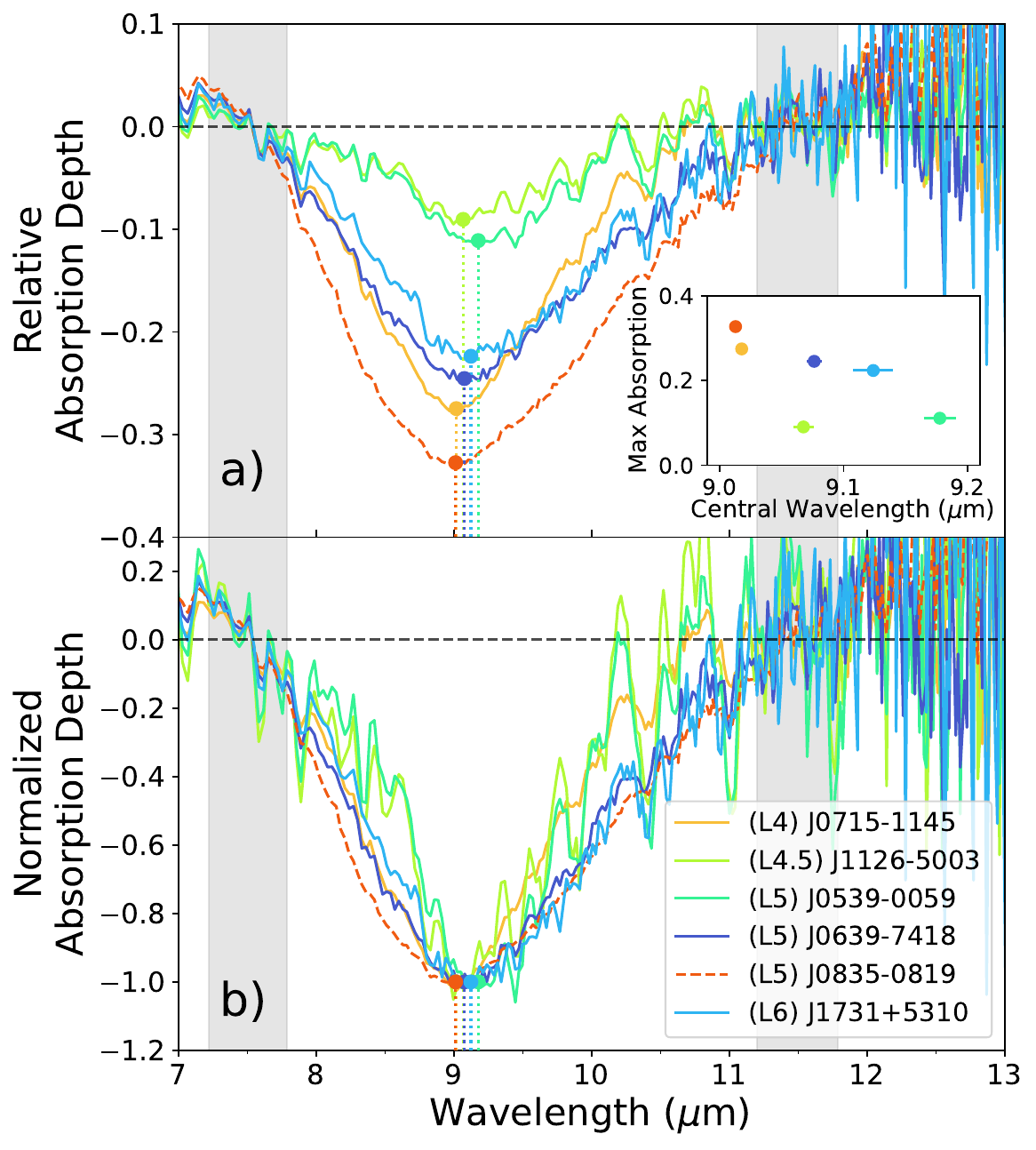}
  \caption{a) Variations of the silicate absorption feature in L4--L6 dwarfs. The relative absorption depth assumes a linear continuum fit to the two regions highlighted in grey, the same as in Fig.~\ref{fig:silicateindexexample}. The central wavelengths of silicate absorption, marked with dotted lines, are estimated from quadratic fits to the 8.6--9.4\,$\umu$m flux. Inset: comparison of maximum relative absorption depth (Equation~\ref{eq:relativeabsorption}) to central absorption wavelength, showing a potential correlation between these values. b) Same as a), but with all profiles normalized to the same maximum absorption.}
  \label{fig:silicateprofiles}
\end{figure}

We can also use the continuum found with Equation \ref{eq:silicateindex} to produce a comparison of continuum-subtracted silicate absorption profiles. While this definition of the continuum may be cruder than dedicated modelling, because existing models do not incorporate condensate opacity, we cannot use such models as an anchor for comparison of dwarfs whose spectra are strongly impacted by this opacity. Fig. \ref{fig:silicateprofiles} shows the absorption profiles after subtracting the continuum. Panel a) shows the relative absorption depth from the continuum for each absorption feature, while panel b) shows all features normalized to the same maximum absorption depth.

Panel a) of Fig. \ref{fig:silicateprofiles} shows clear variations in absorption profile depth in our mid-L dwarfs. The strength of silicate features has been shown by \citet{Suarez2023b} to correlate with viewing geometry, with dwarfs viewed closer to equator-on having stronger silicate features than those viewed closer to pole-on. Inclination data for our observed sample is sparse, with J1126$-$5003 (L4.5) being the only dwarf in this subsample with measured viewing inclination. This dwarf has an inclination angle of 35 $\pm$ 7 degrees \citep{Vos2020}, making it one of the more pole-on dwarfs with known inclinations. Indeed, \textit{Spitzer} IRS data for J1126$-$5003 were included in the correlation by \citet{Suarez2023b}, with its low silicate absorption consistent with its high viewing inclination angle.

We also observe a potential correlation between the relative absorption depth and the wavelength of maximum absorption. To quantify this, we fit a parabola to each absorption profile between 8.6--9.4\,$\umu$m, using the minimum of the fitted parabola as the wavelength of maximum absorption. With the exception of the L4.5 dwarf J1126$-$5003, whose strongest silicate absorption is at a slightly shorter wavelength relative to the rest of the subsample, stronger absorption corresponds to shorter maximum absorption wavelengths. Strongest silicate absorption wavelengths are marked in Fig. \ref{fig:silicateprofiles}, and compared to the maximum absorption depth in the inset of panel a). Several factors are potentially relevant in this relationship, including differences in silicate grain size, variations in Fe/Mg ratio in the silicate grains, or age variations between the mid-L dwarfs \citep{Luna2021, Suarez2023}. Differences in continuum slope are not found to correlate with these potential trends. A larger mid-L sample would be needed to confirm this tentative relationship.

In addition to the variations in depth and central wavelength of the silicate absorption profiles, panel b) of Fig. \ref{fig:silicateprofiles} shows there is also some variation in overall shape. In particular, the profile for the L5 dwarf J0835$-$0819 is the broadest in the subsample, as measured by the equivalent width of the mutually normalized absorption profiles in panel b). This dwarf has a normalized equivalent width that is 47 per cent greater than the next broadest profile (J0639$-$7418, L5), and six times greater than the weakest profile (J1126$-$5003, L4.5). As J0835$-$0819 shows signs of youth \citep{Liu2016}, its particularly broad silicate absorption profile is consistent with the findings of \citet{Suarez2023} who found younger dwarfs to have broader silicate profiles than field dwarfs. 

\subsection{Molecular SiO absorption: discovery and gravity dependency}
\label{sec:sioabsorption}

We observe a shallow absorption feature from $\sim$7.5--9.5 $\umu$m in the L0--L2.5 dwarfs in the left panel of Fig. \ref{fig:spectrallibrary}, marked in grey. The feature is not observed in the L3--L8 dwarfs with silicate absorption in the central column, nor in the T dwarfs. Potential sources for this feature in our 1900--2300 K \citep{Filippazzo2015} L0--L2.5 dwarfs include C$_2$H$_2$, silicate condensates, or SiO. We consider and discard the first two, and settle on SiO as the most likely origin.

C$_2$H$_2$ is known to produce a spectral feature at $\sim$7.7\,$\umu$m in astrophysical contexts \citep[e.g.,][]{Tabone2023}, and has been detected in late-T dwarf atmospheres \citep{Matthews2025}. However, this is a narrow-band feature that does not adequately reproduce the observed broad absorption from $\sim$7.5--9.5\,$\umu$m, with the deepest absorption at $\sim$7.9\,$\umu$m. Various silicate condensates also produce absorption features in this wavelength range \citep{Luna2021}. However, their absorption is centred closer to 9.0\,$\umu$m and does not extend shortwards of $\sim$8.0\,$\umu$m (see Fig. 7 of \citet{Luna2021}).

Molecular SiO offers the best explanation for the observed \hbox{7.5--9.5$\mu$m} absorption. Gas-phase SiO is expected to be present over a wide range of conditions in brown dwarfs \citep{Visscher2010}. It has been detected in emission at 4.0--4.3\,$\umu$m in the dayside spectrum of the ultra-hot Jupiter WASP-121 b \citep{Evans-Soma2025}, whose $\sim$2000 K temperature is similar to that of early-L dwarfs. Ultraviolet 200--300 nm SiO absorption has also been reported in the $\sim$2500--5000~K upper atmosphere of the ultra-hot Jupiter WASP-178 b \citep{Lothringer2022}, at pressure levels similar to those in the atmospheres of K and M giants. Mid-infrared 7.5--9.5~$\mu$m SiO absorption is indeed known to be present in K and M giants \citep{Sloan2015} and in AGB stars \citep{Cami2002, Smolders2012}.

The 7.5--9.5\,$\umu$m gas-phase SiO absorption reported here is the first time such absorption is seen in field L dwarfs. SiO has a band head at $\sim$7.5\,$\umu$m and an extended P-branch tail, consistent with the observed absorption. 
To compare this profile to the observed feature, we remove the continuum in a similar manner as we did for the silicate absorption feature in Section \ref{sec:silicateabsorption} and compare to the molecular SiO opacity from the \citet{Lupu2022} line list at 2200 K and 1 bar, conditions appropriate for early-L dwarfs \citep{Sanghi2023, Luna2021}.

\begin{figure}
  \centering
  \includegraphics[width=\columnwidth, alt={Graph of a spectrum with visible silicon monoxide absorption, with shaded regions, a fitted trendline, and highlighted points showing the method of calculating the silicate index.}]{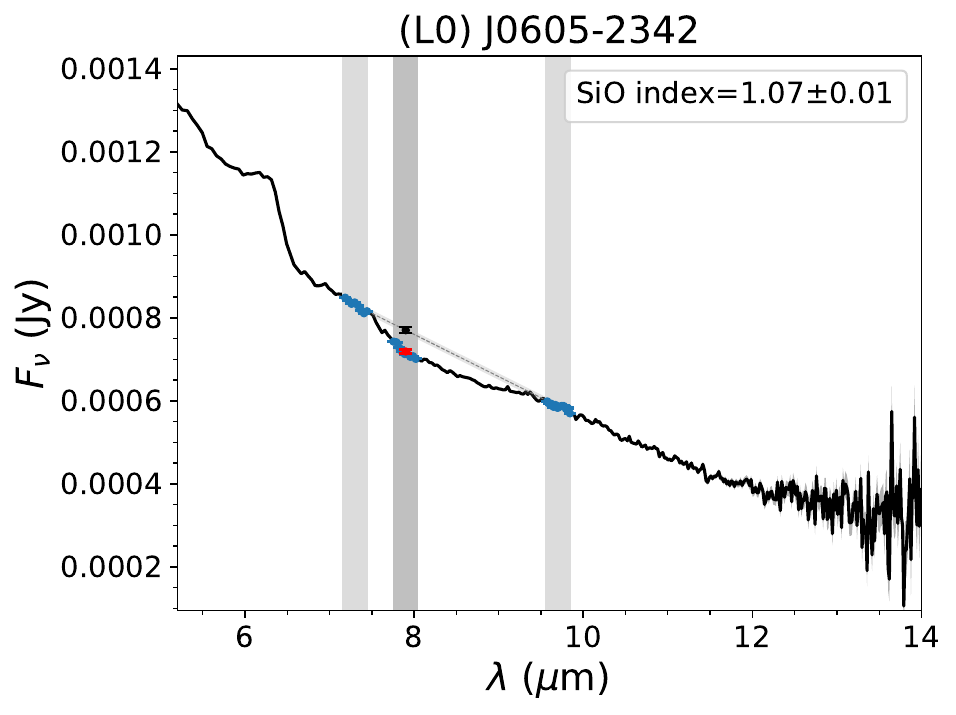}
  \caption{Example of SiO index calculation using the spectrum of the L0 dwarf J0605-2342. The windows used to calculate the index are shaded in grey, with the darker region being the window for absorption and the lighter regions being the windows for continuum fitting. Series of blue data points with thicker error bars delineate the flux values used to calculate the continuum and the mean flux of the absorption region. The SiO index is defined as the ratio of the interpolated continuum flux (black point) to the absorption flux (red point).}
  \label{fig:sioindexexample}
\end{figure}

We define an interpolated continuum and absorption index for the putative SiO feature in a similar manner as the silicate index (Equation \ref{eq:silicateindex}). We fit the linear continuum to two windows of 0.3\,$\umu$m centred at 7.3 and 9.7\,$\umu$m, outside the observed absorption feature. The in-band flux is measured in a window of 0.3\,$\umu$m around 7.9\,$\umu$m, the deepest point of the observed absorption (Fig. \ref{fig:sioindexexample}). 

\begin{equation}
\text{SiO index} = \frac{C_{7.9}}{F_{7.9}}
\label{eq:sioindex}
\end{equation}

The SiO absorption index definition allows us to produce a continuum-subtracted comparison of the absorption features in our early-L dwarfs. We fit a linear continuum to each spectrum as described in Section~\ref{sec:silicateabsorption}, and calculate the relative absorption depth as:

\begin{equation}
\text{Relative absorption depth} = \frac{\text{Observed Flux - Continuum Flux}}{\text{Continuum Flux}}
\label{eq:relativeabsorption}
\end{equation}

\begin{figure}
  \centering
  \includegraphics[width=\columnwidth, alt={Graphs comparing silicon monoxide absorption profiles between early-L brown dwarfs, including an opacity model profile with a similar shape to the observed absorption.}]{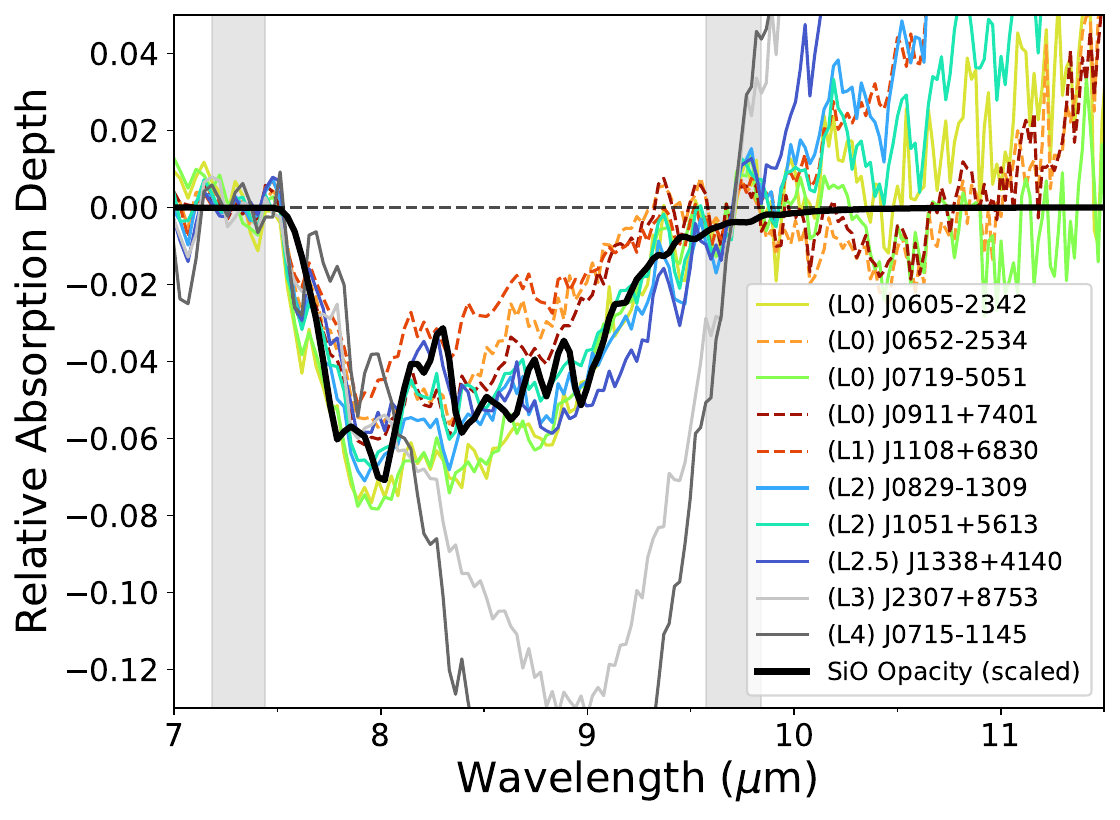}
  \caption{Gas-phase SiO absorption in our L0--L2.5 dwarfs. L3 and L4 dwarf spectra with prominent silicate condensate absorption are shown in grey for comparison. Three L0--L1 dwarfs known or suspected to be young or to have low/intermediate gravity (Section \ref{sec:youngandbinary}) are shown with dashed lines. All of these show marginally weaker SiO absorption than the five bona-fide field L0--L2.5 dwarfs. Continuum fitting regions are highlighted in grey. SiO opacity at 1 bar and 2200 K \citep{Lupu2022} is overplotted, showing consistency with the observed absorption features. The additional narrow absorption features at $\sim$9.4~$\mu$m and $\sim$9.6~$\mu$m are produced by water.}
  \label{fig:sioprofiles}
\end{figure}

Fig. \ref{fig:sioprofiles} shows the relative absorption depths of the SiO feature for the early-L dwarfs in our observed sample. For comparison, we include the spectra of the L3 and L4 dwarfs in our observed sample, subject to the same continuum removal, to show how 8--11 $\umu$m silicate condensate absorption becomes dominant starting at spectral type L3. We also show the molecular SiO opacity from \citet{Lupu2022}, convolved to the same resolution as MIRI LRS and overplotted in black. To approximately scale this opacity to the observed absorption, we calculate the equivalent width of the SiO absorption feature for each of our early-L dwarfs. We then take the mean of these equivalent widths to represent the typical integrated absorption depth of the SiO feature across our early-L dwarfs, and scale the opacity profile to the same equivalent width.

A more accurate comparison would require integrating this opacity over the full range of an atmospheric model. However, even this empirical approximation shows that the overall shape of the SiO opacity, as well as specific features such as the broader red wing and reduced absorption at $\sim$8.2\,$\umu$m, are consistent with all observed early-L spectra. This confirms that this feature is indeed produced by SiO gas.

Fig. \ref{fig:sioprofiles} also shows a potential dependence of absorption with surface gravity, where all three of the young/low-gravity objects as discussed in Section \ref{sec:youngandbinary} exhibit weaker SiO absorption than any of the four higher-gravity dwarfs. 
Other than the small range in absorption depth, the SiO feature appears broadly similar among the L0--L2.5 dwarfs in this subsample.

\begin{figure}
  \centering
  \includegraphics[width=\columnwidth, alt={Graphs comparing silicon monoxide absorption profiles between early-L brown dwarfs observed with James Webb and late-M red dwarfs observed with Spitzer.}]{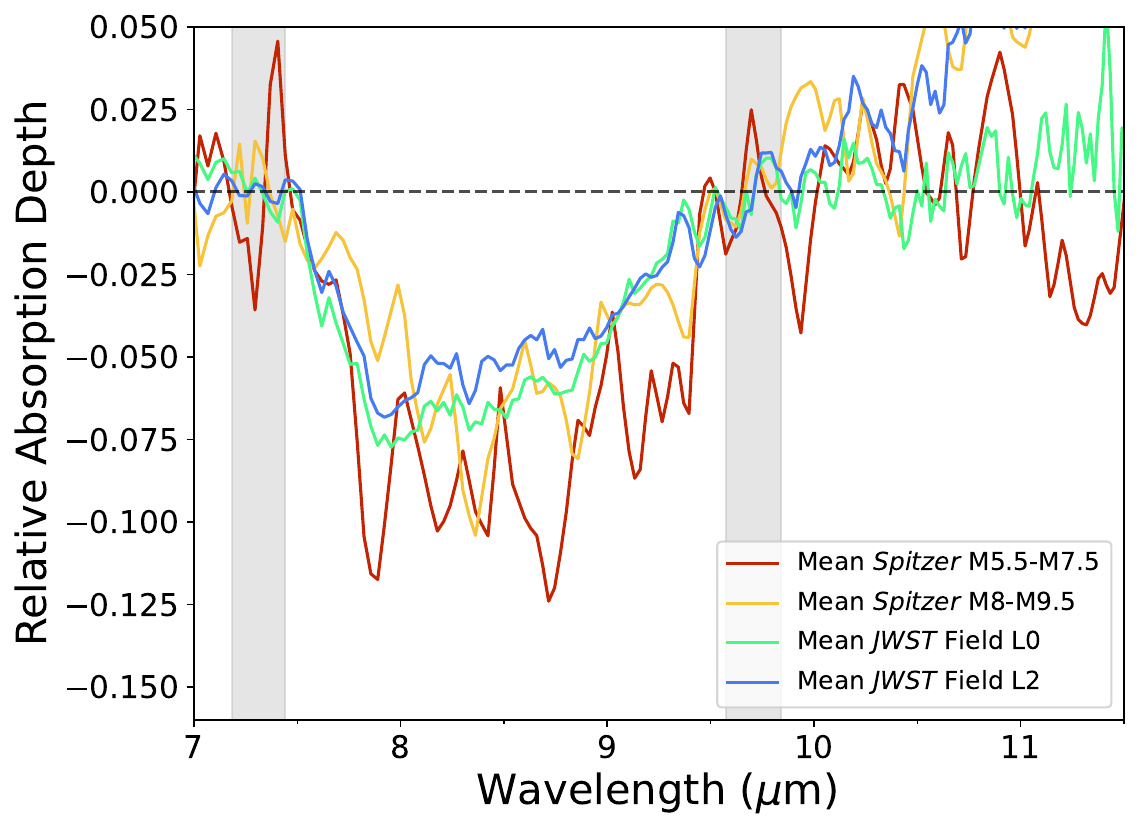}
  \caption{Comparison of relative absorption  for the binned spectra of M5.5--M7.5 and M8--M9 dwarfs (\textit{Spitzer} IRS) as well as L0 and L2 dwarfs (MIRI LRS). SiO absorption is visible in \textit{Spitzer} spectra of late-M dwarfs.}
  \label{fig:siowithspitzer}
\end{figure}

As this SiO absorption is observed in the earliest-type L dwarfs in our observed sample, it is then logical to consider if it may be present in late-M dwarfs as well. For this, we turn to the \textit{Spitzer} IRS spectra of late-M dwarfs from \citet{Suarez2022}. We select the spectra with the lowest noise levels, and fit continua to the SiO feature and find the relative fluxes in the same manner as with the \textit{JWST} spectra. We exclude only the high-SNR \textit{Spitzer} spectrum of the M5 dwarf J2238$-$1517 (EZ Aqr) as it is an unresolved triple \citep{Delfosse1999}. We then bin the resulting relative absorption spectra into inverse-variance weighted mean M5.5--M7.5 and M8--M9 residual spectra. The same binning is performed on the relative absorption spectra of the two field L0 and two field L2 dwarfs in our \textit{JWST} sample. These mean absorption spectra are compared in Fig. \ref{fig:siowithspitzer}.

While the \textit{Spitzer} IRS spectra are noisier than our MIRI LRS spectra, the presence of absorption similar in extent and depth to the SiO absorption observed in our L dwarfs can also be seen in these M dwarf spectra. We thus conclude that gas-phase SiO absorption occurs in dwarfs as early as type M5.5, with strengths similar to these seen in early L dwarfs. However, the SiO absorption in M5.5--M9.5 dwarfs reported here is significantly weaker than in M giants \citep{Sloan2015}, contrary to the potential dependence on surface gravity in L0--L2.5 dwarfs seen in Fig.~\ref{fig:sioprofiles}.

The SiO indices calculated with Equation \ref{eq:sioindex} are shown for both our observed sample and the \textit{Spitzer} IRS data from \citet{Suarez2022} in the bottom panel of Fig. \ref{fig:silicateindices}. We caution against using this index and continuum definition at spectral types later than $\sim$L2, as the early-L gas-phase SiO absorption feature begins to blend into the main mid-L silicate absorption feature. Disentangling these two features would require dedicated modelling  that could reveal continued SiO absorption in dwarfs cooler than early-L.

\section{Spectroscopic Diversity: Key Inferences}
\label{sec:withintypecomparison}

Our observed sample of 23 L0--T6 dwarfs contains three groups of objects with similar spectral sub-types: eight early L dwarfs (L0--L2.5), six mid-L dwarfs (L4--L6), and six T dwarfs (T1.5--T6). We explore each of these in Sections~\ref{sec:L0comparison}--\ref{sec:T1-T6comparison} to assess spectroscopic diversity under similar atmospheric conditions. 

\subsection{L0--L2.5 dwarfs: SiO and the onset of silicate absorption}
\label{sec:L0comparison}

In Fig. \ref{fig:L0comparison}, we compare our eight L0--L2.5 dwarf spectra. Panel a) shows the spectra scaled to a distance of 10 pc, using the parallaxes from Table \ref{table:observations}. To show the subtle variations in spectral shape across the L0--L2.5 spectral type range, panel b) shows the spectra normalized to the average flux between 6.1\,$\umu$m and 6.3\,$\umu$m, with the mean spectrum of the two field-aged L0 dwarfs J0605$-$2342 and J0719$-$5051 then subtracted from each normalized spectrum. This reference spectrum is comprised of the two early-L dwarfs with the weakest silicate absorption between 8--11~$\mu$m. We show uncertainties as a single encircled error bar at the short wavelength end of each spectrum. The high SNR of our MIRI LRS spectra at $<$10 $\mu$m means any uncertainty in the absolute flux is dominated by parallax uncertainties. For our L0--L2.5 dwarfs, the combined flux and parallax uncertainties at $<$10$\micron$ are negligible.

\subsubsection{Overluminosity due to binarity and youth}
\label{sec:L0overluminosity}

\begin{figure}
  \centering
  \includegraphics[width=\columnwidth, alt={Graphs comparing early-L dwarf spectra, showing stratification in luminosity and smooth progression in both silicate and silicon monoxide absorption.}]{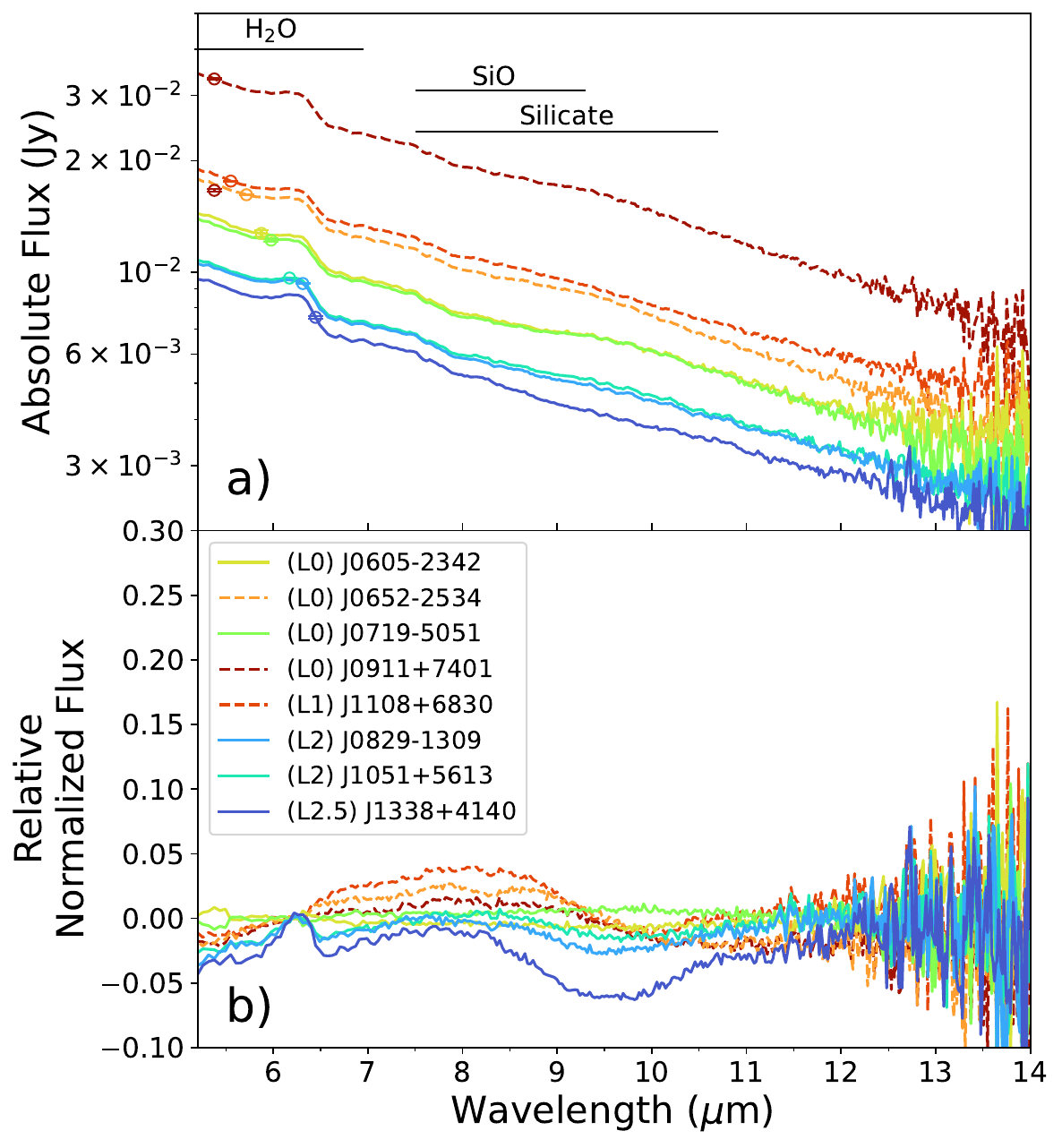}
  \caption{a) Comparison of distance-scaled L0--L2.5 dwarf spectra. Dwarfs with published indications of youth are shown with dashed lines. b) Comparison of L0--L2.5 dwarf spectra normalized between 6.1--6.3\,$\umu$m with the average spectrum of field-aged L0 dwarfs J0605$-$2342 and J0719$-$5051 then subtracted. Silicate absorption is seen as a broad depression between 8$-$11 $\umu$m as early as spectral type L0, with varying degrees of silicate absorption between our L0 dwarfs.}
  \label{fig:L0comparison}
\end{figure}

Panel a) of Fig. \ref{fig:L0comparison} shows a clear distinction between dwarfs of different ages, spectral types, and binarity status, with the candidate binary L0 dwarf J0911+7401 being the brightest across the 5--14 $\umu$m range. Also, the fluxes of the potentially young (Section \ref{sec:youngandbinary}) J0652$-$2534 (L0) and J1108+6830 (L1) are systematically $\sim$20 per cent brighter when compared to the remaining two likely field L0 dwarfs. The shared overluminosity of these two dwarfs is consistent with expectations for younger brown dwarfs being overluminous at a given temperature \citep{Faherty2016}.

We look to evolutionary models of L dwarfs from BT-Settl \citep{Allard2014} to examine the predicted effects of youth on luminosity, assuming the nominal \citet{Sanghi2023} temperatures of 2100--2300 K for L0--L1 dwarfs. Using the predicted luminosity from these evolutionary models, we find {that the $\sim$20\% overluminosity of J0652$-$2534 (L0) and J1108+6830 (L1) suggests ages of about 10--300~Myr for these two objects.}

\subsubsection{Differentiation between molecular SiO and silicate}
\label{sec:L0sioandsilicate}

Panel b) of Fig. \ref{fig:L0comparison} shows absorption beyond 8.5\,$\umu$m at varying strengths in nearly all our L0--L2.5 spectra, which we attribute to silicate condensate absorption known from later spectral types. The onset of silicate absorption at spectral type L0 is consistent with the L1 onset inferred from lower-SNR \textit{Spitzer} IRS spectra in \citet{Suarez2023}. Silicate absorption may be present even in late-M dwarf spectra. However, higher-SNR M dwarf spectra than currently exist from \textit{Spitzer} IRS would be needed to assess this.

The relative fluxes of the three dwarfs with indications of youth discussed in Section \ref{sec:youngandbinary} as shown in panel b) show a region of elevated flux between 7.5--9.0\,$\umu$m, as expected due to the reduced SiO absorption in these lower-gravity dwarfs compared to the rest of the L0--L2.5 subsample (Fig. \ref{fig:sioprofiles}). Panel b) of Fig. \ref{fig:L0comparison} shows that the SiO and silicate absorption features occur in dwarfs of the same spectral types, with overlapping wavelength ranges that present challenges to disentangling their effects. Maximizing the diagnostic utility of the SiO feature will require accurate deblending of this absorption feature from silicate condensate absorption.

\subsection{L4--L6 dwarfs: wide diversity in silicate strength and anti-correlation with water absorption}
\label{sec:L4-L6comparison}

\begin{figure}
  \centering
  \includegraphics[width=\columnwidth, alt={Graphs comparing mid-L dwarf spectra, with water and silicate absorption features highlighted to emphasize their anticorrelation.}]{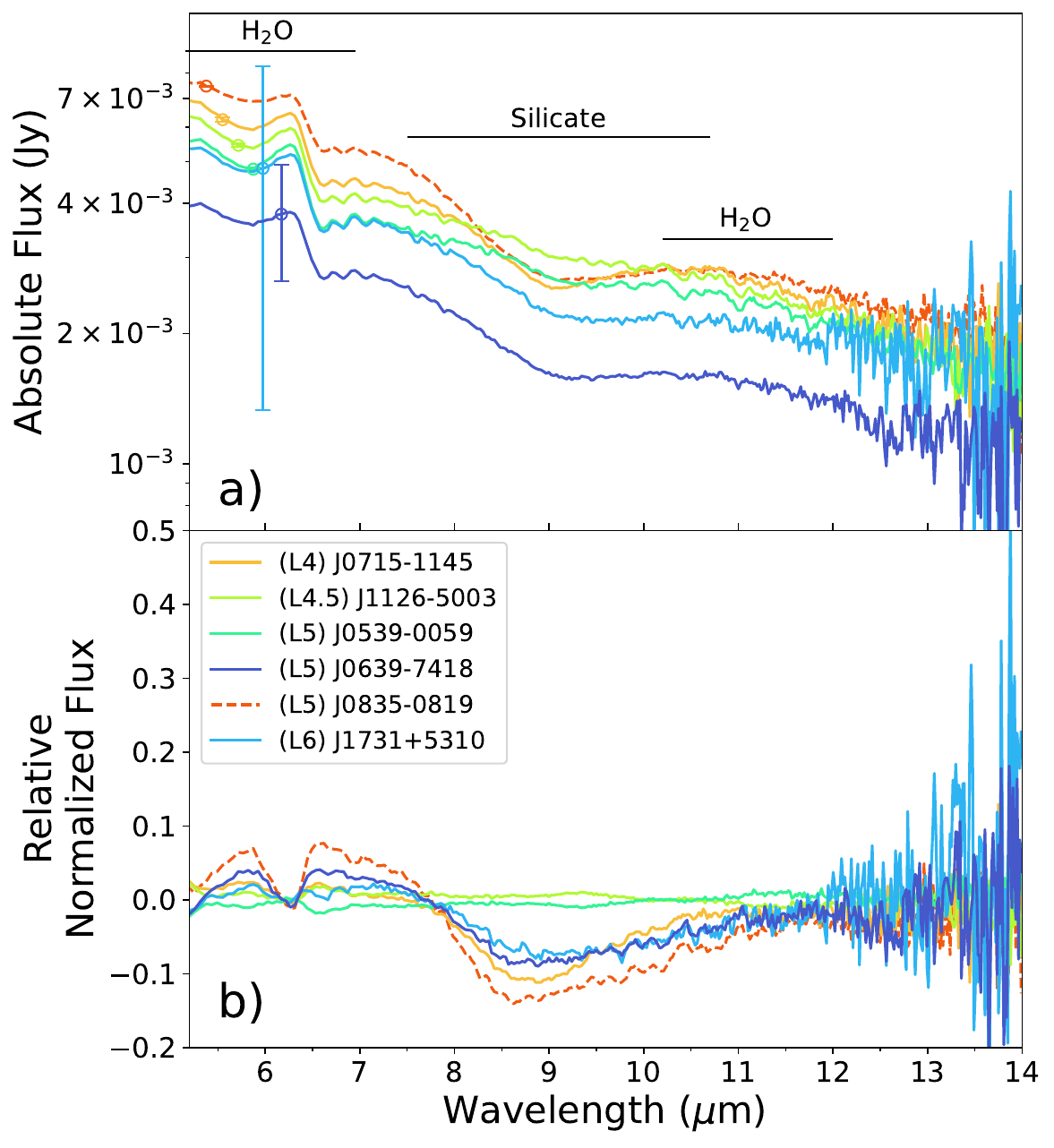}
  \caption{a) Comparison of distance-scaled L4--L6 dwarf spectra. The L5 dwarf with published indications of youth is shown with a dashed line. The large uncertainty ranges for J0639$-$7418 and J1731+5310 are due to their less accurate parallaxes. b) Comparison of L4--L6 dwarf spectra normalized between 6.1--6.3\,$\umu$m with the spectrum of low-silicate absorption dwarfs J1126$-$5003 (L4.5) and J0539$-$0059 (L5) then subtracted. Water and silicate absorption depths appear anticorrelated, consistent with expectations of silicate clouds obscuring water absorpton \citep{Suarez2023}.}
  \label{fig:L4-L6comparison}
\end{figure}

Fig. \ref{fig:L4-L6comparison} compares our six L4--L6 dwarf spectra in a similar manner to Fig. \ref{fig:L0comparison}. Panel a) shows the spectra scaled to a 10 pc distance, while panel b) shows the spectra normalized between 6.1 and 6.3\,$\umu$m, with the mean spectrum of low-silicate absorption dwarfs J1126$-$5003 (L4.5) and J0539$-$0059 (L5) then subtracted from each normalized spectrum. Significant uncertainty in the flux of J0639$-$7418 (L5) and J1731+5310 (L6) is apparent in panel a), which arises due to the less tightly constrained distances for these dwarfs compared to the other mid-L dwarfs \citep[][]{Smart2018, Best2025}. The L5 dwarf J0835$-$0819 is slightly overluminous relative to the other mid-L dwarfs. As with the L0 dwarf J0652$-$2534, \citet{Sanghi2023} find indications of intermediate gravity for J0835$-$0819, suggesting this overluminosity may be a result of youth.

Panel b) shows that the depth of the silicate absorption band mirrors that seen in the main 5--7\,$\umu$m water absorption features in these spectra. Positive excesses in the residuals on either side of the 6.1--6.3~$\mu$m normalization region result from weaker water absorption compared to the mean spectrum of the two low-silicate dwarfs J1126$-$5003 (L4.5) and J0539$-$0059 (L5). The L5 dwarfs J0835$-$0819 and J0639$-$7418 thus show the weakest water absorption, correlating with strong silicate absorption in both objects. At the same time, the strongest water absorption is observed in the two low-silicate dwarfs: J1126$-$5003 (L4.5) and J0539$-$0059 (L5). As mentioned in Section \ref{sec:L0comparison}, such anticorrelation between water and silicate absorption is consistent with the discussion in \citet{Suarez2023} of silicate clouds obscuring molecular absorption features. 

\subsection{T1.5--T6 dwarfs: more possible binaries, and possible detection of CS$_2$}
\label{sec:T1-T6comparison}

\begin{figure}
  \centering
  \includegraphics[width=\columnwidth, alt={Graphs comparing T dwarf spectra, showing stratification in luminosity and absorption by methane and ammonia.}]{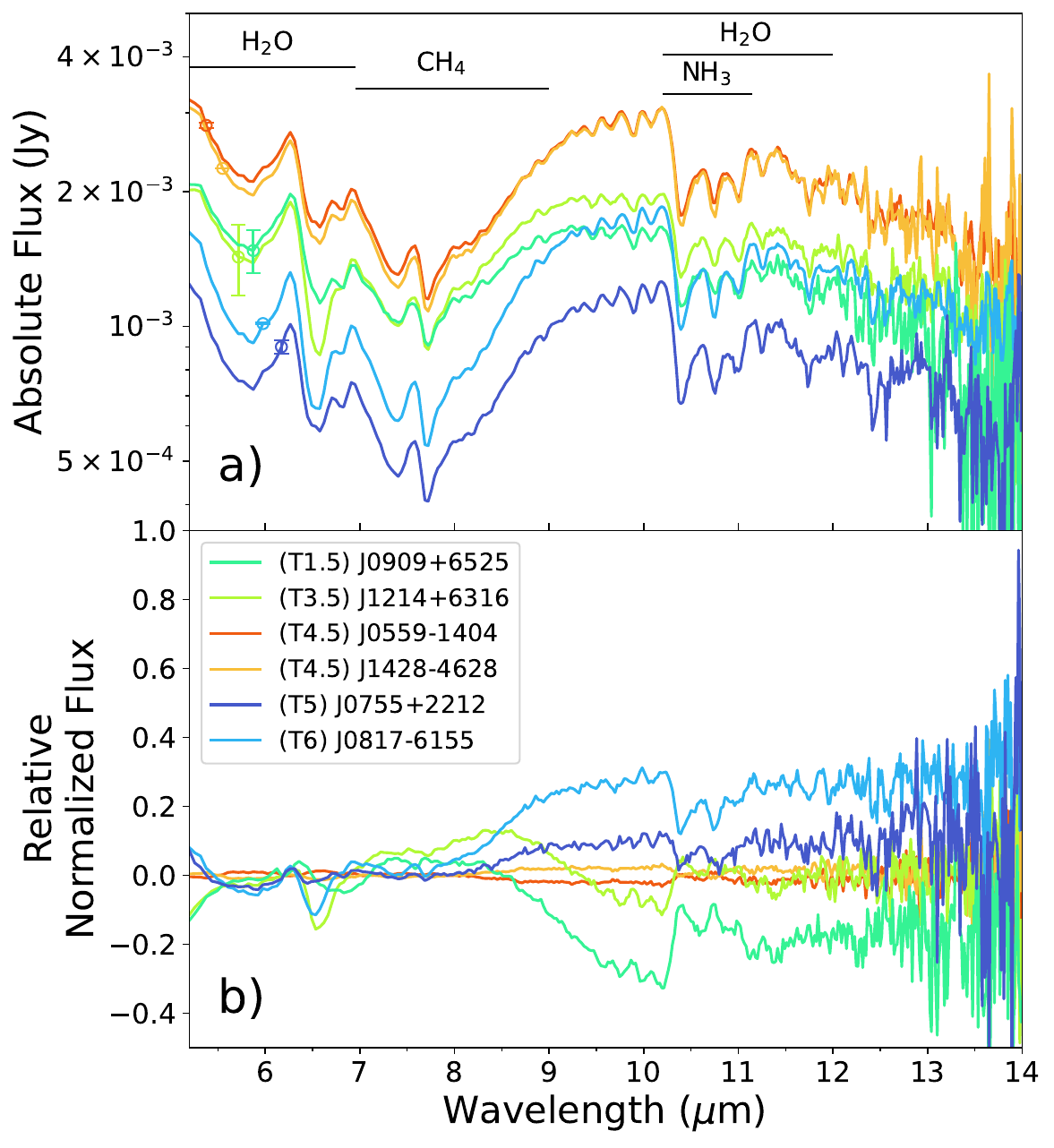}
  \caption{a) Comparison of distance-scaled T1.5--T6 dwarf spectra. b) Comparison of T4--T6 dwarf spectra normalized between 6.1--6.3\,$\umu$m with the spectrum of T4.5 dwarfs J0559$-$1404 and J1428$-$4628 then subtracted. An unidentified absorption feature centred at $\sim$6.6 $\umu$m is visible in the spectrum of T3.5 dwarf J1214+6316 in panel b), potentially corresponding to CS$_2$ absorption.}
  \label{fig:T1-T6comparison}
\end{figure}

The six T dwarfs in our observed sample range in spectral type from T1.5 to T6 and are shown in Fig. \ref{fig:T1-T6comparison}. As with the L0 and L4--L6 dwarfs, panel a) shows the spectra scaled to an absolute flux at a 10 pc distance, and panel b) shows the spectra normalized between 6.1\,$\umu$m and 6.3\,$\umu$m with the mean spectrum of the T4.5 dwarfs J0559$-$1404 and J1428$-$4628 subtracted from each normalized spectrum. Unlike the L0--L2.5 and L4--L6 sequences in Fig.s~\ref{fig:L0comparison} and ~\ref{fig:L4-L6comparison}, the absolute flux sequence of the T1.5--T6 is not monotonic with spectral type, with binarity being the main culprit.

\subsubsection{Confirmed and possible unresolved binarity}
\label{sec:Tbinaries}

The T4.5 dwarf J1428$-$4628 is discussed as a newly resolved binary in Section~\ref{sec:resolvedbinaries}, and
the T4.5 dwarf J0559$-$1404 as an overluminous candidate unresolved binary in Section~\ref{sec:unresolvedbinaries}. Their spectra in Fig.~\ref{fig:T1-T6comparison} a) are spectrophotometrically very similar, validating the candidate binarity of J0559$-$1404 over the full near- to mid-IR wavelength range. We note two additional unresolved potential binaries that we consider as only possible, rather than confirmed or candidate binaries: the T6 dwarf J0817$-$6155 (discussed below) and the T3.5 dwarf J1214+6316 (discussed in Section~\ref{sec:6.6umabsorption}).

The T6 dwarf J0817$-$6155 is systematically $\sim$50 per cent brighter compared to the T5 dwarf J0755+2212. The reason for this apparent overluminosity is unclear. J0817$-$6155 was identified by \citet{Cheng2025} as a potential intermediate-gravity dwarf through comparison with SpeX $R\sim100$ template spectra, suggesting possible youth. However, higher-resolution spectra analysed in \citet[][$R\sim1200$]{Artigau2010} and \citet[][$R\sim40,000$]{Tannock2022} consistently found $\log g$ values of 5.0 to 5.1 $\pm$ 0.1, inconsistent with youth for a T dwarf. \citet{Cheng2025} also note a slightly elevated \textit{Gaia} RUWE value of 1.24 for J0817$-$6155 (T6), which may suggest unresolved binarity. While this dwarf is not resolved in our target acquisition imaging (Section \ref{sec:resolvedbinaries}), it is one of two dwarfs in our sample observed with the wide-PSF F1500W filter due to brightness constraints, limiting the spatial resolving power of our target acquisition imaging.

J0817$-$6155 is thus a possible $<$0.5 arcsec unequal-flux binary. However, the absolute flux discrepancy is within the uncertainty band for spectral type vs.\ absolute magnitude relationships (see Fig.~\ref{fig:absmagvsspectraltype}), and the evidence for unresolved binarity is weaker than for our candidate equal-flux unresolved binaries (Section~\ref{sec:unresolvedbinaries}). 

\subsubsection{Potential CS$_2$ absorption in a T3.5 dwarf}
\label{sec:6.6umabsorption}

The 6.5~$\mu$m water absorption band in the T3.5 dwarf J1214+6316 appears unusually strong for this dwarf's spectral subtype. Other absorption features, including the $\geq$10.3~$\mu$m NH$_3$ band, show absorption increasing monotonically in strength with spectral subtype, as seen in the residuals with respect to the mean T4.5 spectrum in Fig.~\ref{fig:T1-T6comparison} b). However, the same monotonic sequence is not observed in the 6.5~$\mu$m residuals, where the T3.5 dwarf J1214+6316 is an outlier with unusually strong 6.5~$\mu$m water absorption, stronger than even in the coolest (T6) dwarf J0817$-$6155.

\begin{figure}
  \centering
  \includegraphics[width=\columnwidth, alt={Graph showing the scaled and added spectra of two observed sample brown dwarfs, compared to a target spectrum, with a relatively flat residual except one unidentified absorption feature overlaid with the absorption profile of carbon disulphide.}]{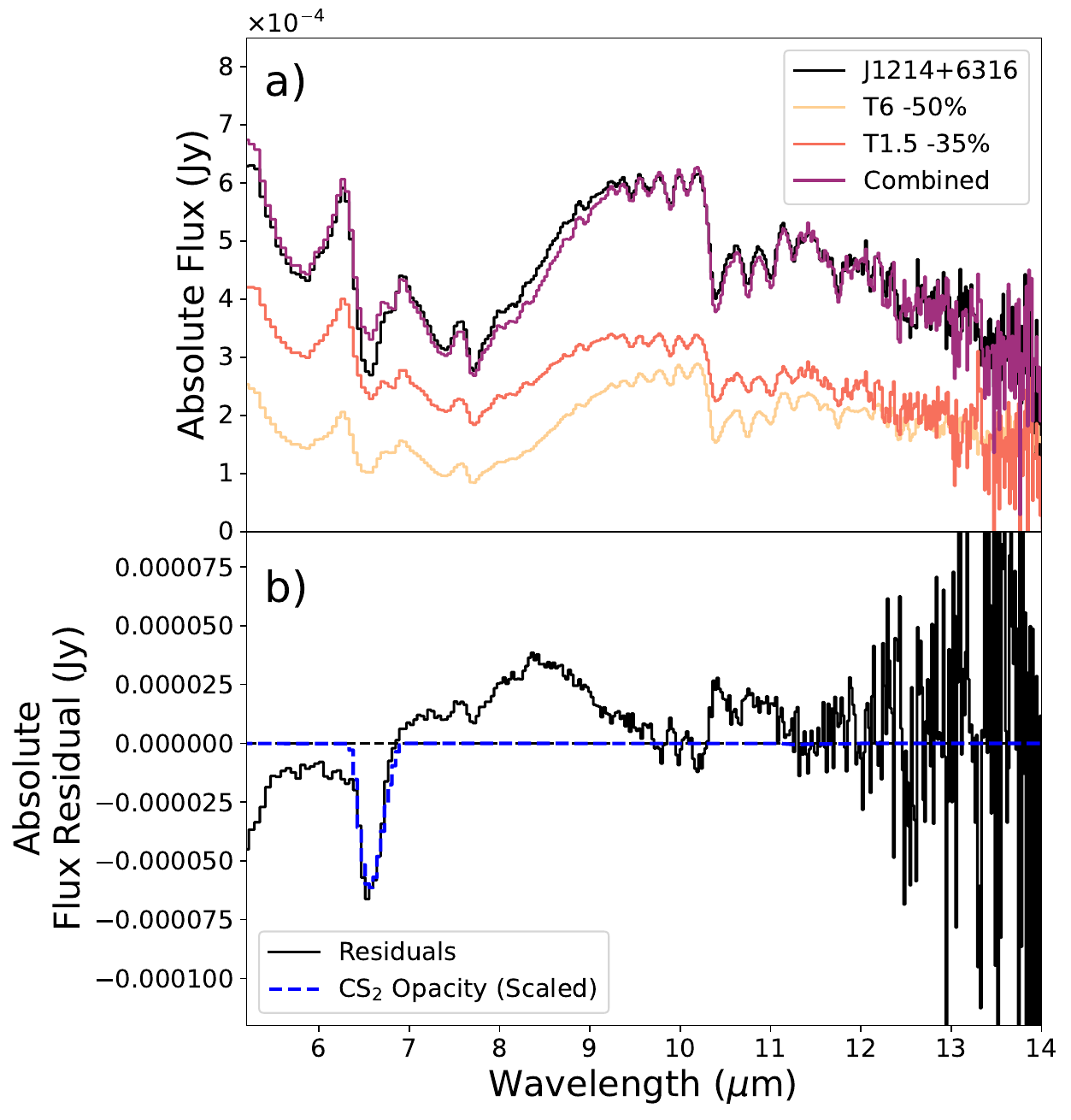}
  \caption{a) Best-fitting combination of our T1.5 and T6 dwarf spectra to the spectrum of T3.5 dwarf J1214+6316. The template spectra are scaled down in flux to match the absolute flux of the T3.5 dwarf. The combination generally produces a strong fit, except for an unidentified absorption feature centred at $\sim$6.6 $\umu$m. b) Residuals from the binary fit, with the scaled absorption profile of CS$_2$ overlaid in blue. The unidentified absorption feature at $\sim$6.6 $\umu$m is well-fit by CS$_2$ absorption.}
  \label{fig:T3binary}
\end{figure}

The unusually deep 6.5~$\mu$m absorption in the T3.5 dwarf J1214+6316 could be an indication of a cooler unresolved companion, given that 6.5~$\mu$m absorption is strongest in the T6 dwarf. Indeed, J1214+6316 was identified as a possible T2+T6 spectral binary from low-resolution NIR spectra by \citet{Geissler2011}. We confirm that a linear combination of our T1.5 and T6 mid-IR spectra produces a close match to the spectrum of J1214+6316 if the fluxes of the components are scaled down by 35\% and 50\% respectively (Fig.~\ref{fig:T3binary}). The need to scale both template spectra down by nearly half suggests that J1214+6316 is not an overluminous candidate for unresolved binarity. Furthermore, archival high-resolution imaging with Keck AO, as well as lower-resolution \textit{Spitzer} IRAC imaging from the \textit{Spitzer} Heritage Archive and our target acquisition imaging with MIRI, all fail to show any visible binarity at projected angular separations of $\gtrsim$0.05 arcsec. However, we note that the parallax for this dwarf is poorly constrained (9\% uncertainty; Table~\ref{table:observations}) compared to others ($<$1\% uncertainties) in our observed sample. If the measured parallax is greater than the true value at the 2$\sigma$ level, the required scaling of the component fluxes becomes marginal. We thus suggest this dwarf as a possible, though not necessarily likely, unresolved binary, similar to the T6 dwarf J0817$-$6155.

The comparison of the T3.5 dwarf J1214+6316 to a linear combination of a T1.5 and a T6 dwarf in Fig.~\ref{fig:T3binary} reveals a strong residual near the 6.5~$\mu$m water feature that is offset to a slightly longer wavelength, $\sim$6.6$\mu$m. The absorption trough in J1214+6316 is indeed slightly redward of the water absorption in the T6 dwarf in Fig.~\ref{fig:T1-T6comparison} b). Fitting a Gaussian to the unidentified absorption {residual}, we find it to be centred at 6.58 $\umu$m (1520 cm$^{-1}$). Notably, this closely matches the  $\nu_3$ absorption band of carbon disulphide \citep{Plyler1947}. As with the SiO absorption in Section \ref{sec:sioabsorption}, we compare the opacity of CS$_2$ from an external line list to the observed residual. Here, we use theoretical opacities calculated using the HAPI package \citep{Kochanov2016} based on the HITRAN molecular opacities database \citep{Gordon2026}. We calculate the opacity of CS$_2$ at a temperature of 1100 K and pressure of 1 bar, appropriate for early-T dwarfs \citep{Sanghi2023}. This opacity is convolved to the resolution of MIRI LRS and scaled to match the depth of the observed residual. The result is plotted in blue in panel b) of Fig. \ref{fig:T3binary}, and shows the similarity in absorption profile between CS$_2$ and our observed absorption feature, including a sharper blue wing and broader red wing.

Absorption from CS$_2$ has been observed in the atmospheres of Jupiter \citep{Atreya1995}, Venus \citep{Mahieux2023}, and exoplanets WASP-80 b \citep{Triantafillides2026} and (tentatively) TOI-270 d \citep{Felix2025}. However, these detections were of shorter-wavelength bands at 0.2 $\umu$m, 3.5 $\umu$m, and 4.6 $\umu$m respectively, and in all these cases the presence of CS$_2$ is explained as the result of external factors: shock-induced chemistry following the Shomeaker-Levy 9 impact on Jupiter, or UV irradiation-assisted photochemistry on Venus or close-in exoplanets. No such external factors are anticipated or evident for the T3.5 dwarf J1214+6316, unless one of the putative components in this potential T1.5+T6 binary is a source of strong (e.g., auroral) UV radiation. Hence, the association of the $\sim$6.6~$\mu$m excess absorption with CS$_2$ is at present only a conjecture, requiring follow-up spectroscopy for confirmation. If confirmed, this would be the first detection of carbon disulphide in a brown dwarf.

\section{Underluminosity of single, field L and T dwarfs.}
\label{sec:absolutemagnituderelations}

Our sample selection avoided known binaries or young dwarfs, although dwarfs with unconfimed binarity or youth signatures were allowed. With the combined leverage of accurate \textit{JWST} spectrophotometry and \textit{Gaia} parallaxes, we have now been able to validate previously suggested binarity or youth in our observed sample. We have confirmed two resolved and two candidate binaries, and four additional dwarfs that are young (Section~\ref{sec:youngandbinary}).

\begin{figure*}
  \includegraphics[width=0.85\paperwidth, alt={Graphs comparing luminosity as a function of spectral type to empirical relations, including corresponding residuals for each band.}]{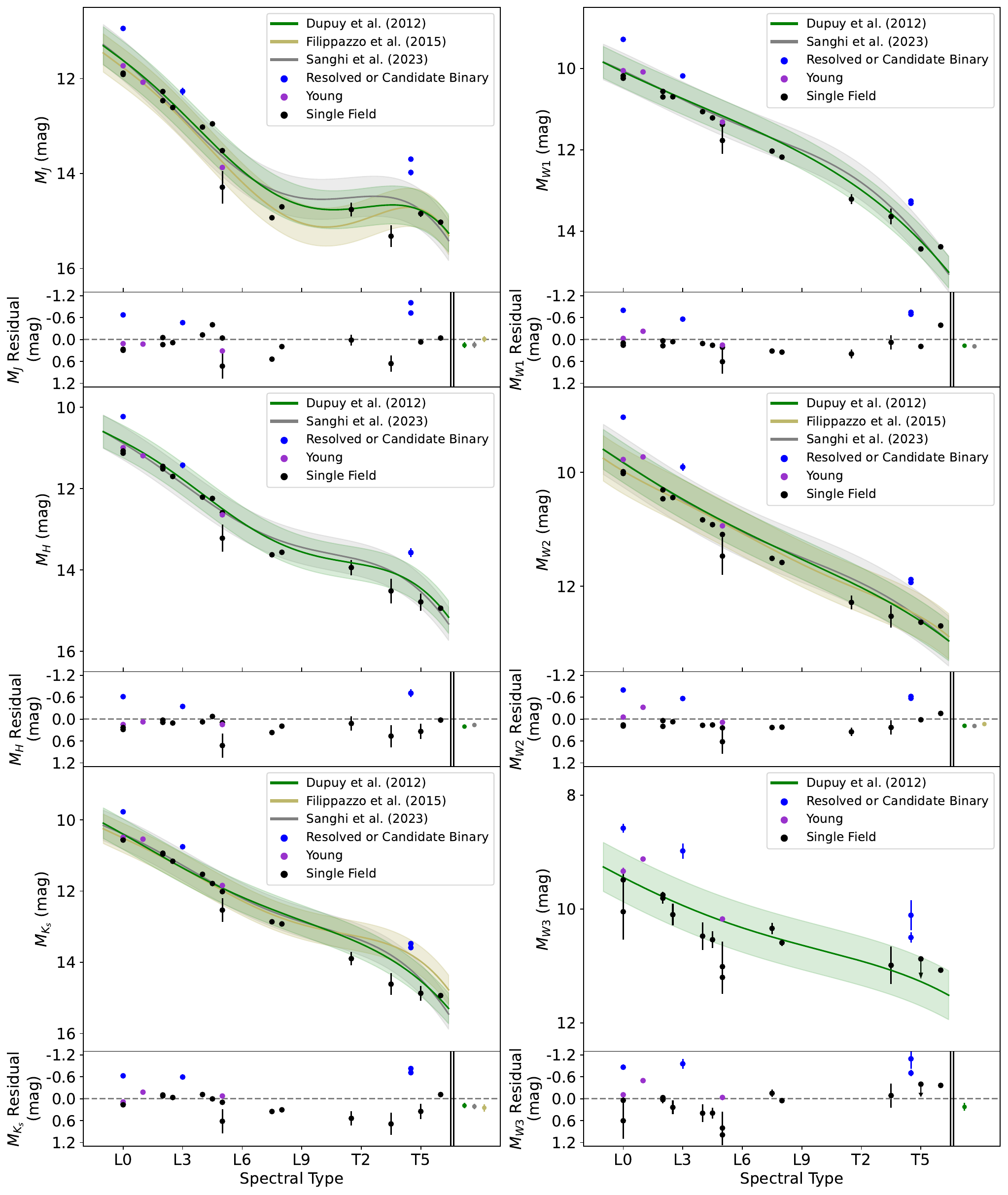}
  \caption{\textit{2MASS} \textit{J}, \textit{H}, and \textit{$K_s$} and \textit{WISE} W1, W2, and W3 band absolute magnitude versus spectral type for the dwarfs in our observed sample. The field-age empirical relations from \citet{Dupuy2012}, \citet{Filippazzo2015}, and \citet{Sanghi2023} along with their observed scatter are overplotted as coloured lines with 1$\sigma$ uncertainty bands. Residuals are shown relative to the \citet{Dupuy2012} relations for each photometric band. Mean residuals for field dwarfs in our observed sample are also shown to the right of the full set of residuals, with colours corresponding to each relation. Resolved or candidate binaries are shown in blue, while dwarfs with probable low gravity are shown in purple.}
  \label{fig:absmagvsspectraltype}
\end{figure*}

The remaining sample of presumed single, field-aged L and T dwarfs can be compared against established empirical relations of absolute magnitude (\textit{2MASS} \textit{J}, \textit{H}, and \textit{$K_s$} and \textit{WISE} W1, W2, and W3) vs.\ spectral type \citep{Dupuy2012, Filippazzo2015, Sanghi2023}. These relations were designed based on samples screened to remove known binaries, subdwarfs, and young/very low gravity sources, so should be comparable to our presumed-single field sample. We present this analysis in Fig.~\ref{fig:absmagvsspectraltype}. Newly resolved or candidate binaries are highlighted in blue, and low gravity dwarfs are highlighted in purple. The two additional possible T dwarf unequal-flux binaries (Sections~\ref{sec:youngandbinary}, \ref{sec:Tbinaries}) have been assumed single, but as we discuss below, the comparison does not hinge on their inclusion. For each of the photometric bands, we show residuals relative to the \citet{Dupuy2012} absolute magnitude vs.\ spectral type relations, as this is the only source with published relations for all six bands analysed here. We also show the mean absolute magnitude residual for our field single dwarfs compared to each relation in the bottom right corner of the panel for each photometric band in Fig. \ref{fig:absmagvsspectraltype} and list the results in Table \ref{table:residuals}.
The W3 photometry for J0909+6525 (T1.5) is contaminated by a nearby source, so we have omitted it from this Fig. and the calculations. We have also removed the L6 dwarf J1731+5310 which does not have a measured parallax.

\begin{table}
    \caption{Mean absolute magnitude residuals for single field dwarfs with good photometry in our observed sample compared to empirical relations. $>3\sigma$ residuals are bolded.}
    \centering
    \begin{tabular}{|c|c c c|}
        \hline
        \hline
        Band & Dupuy \& Liu & Filippazzo & Sanghi et al. \\
        & (2012) & et al. (2015) & (2023) \\
        \hline
        \textbf{\textit{J}} & 0.155 $\pm$ 0.079 & -0.013 $\pm$ 0.079 & 0.146 $\pm$ 0.092 \\
        \textbf{\textit{H}} & \textbf{0.204 $\pm$ 0.053} & $-$ & 0.161 $\pm$ 0.062 \\
        \textbf{\textit{K$_s$}} & 0.187 $\pm$ 0.071 & 0.247 $\pm$ 0.101 & 0.212 $\pm$ 0.075 \\
        \textbf{W1} & \textbf{0.168 $\pm$ 0.056} & $-$ & \textbf{0.186 $\pm$ 0.061} \\
        \textbf{W2} & \textbf{0.182 $\pm$ 0.044} & \textbf{0.135 $\pm$ 0.041} & \textbf{0.187 $\pm$ 0.044} \\
        \textbf{W3} & 0.224 $\pm$ 0.110 & $-$ & $-$ \\
        \hline
    \end{tabular}
    \label{table:residuals}
\end{table}

The sample of 13 remaining L0--T6 dwarfs show tentative, but nearly ubiquitious 1--4$\sigma$ underluminosity throughout the NIR and mid-IR, compared to all three empirical relations (Table~\ref{table:residuals}). Notably, the observed sample is as a whole consistently $>$3$\sigma$ underluminous in both W1 and W2. Only the $J$-band \citet{Filippazzo2015} absolute magnitude vs.\ spectral type relationship is on average accurate. Removing the additional two possible but unconfirmed binaries--J1214+6316 (T3.5) and J0817$-$6155 (T6)--increases the significance of the underluminosity in the \textit{WISE} bands by an average of $\sim$1$\sigma$, while the NIR results remain consistent to $<$0.5$\sigma$.

The observed underluminosity is not constant with spectral type. The NIR relations on Fig. \ref{fig:absmagvsspectraltype} are more accurate for the seven field L0--L6 dwarfs in our observed sample, while the six field L7.5--T5 dwarfs are systematically fainter than expected. The systematic offset in the mid-IR W1 and W2 bands is more uniform across our observed sample. The increase of the NIR offset with spectral type compared to the more consistent mid-IR underluminosity could suggest contamination by undetected T and Y dwarf companions, which would more strongly contribute to the flux near 3--5 $\mu$m (\textit{W1} and \textit{W2}).

The true field dwarf absolute magnitude-spectral type relations are thus potentially both 0.15--0.20 mag fainter and lower in scatter than the three relations referenced here (Fig. \ref{fig:absmagvsspectraltype}). Given the small size of our observed sample of field single L0--T6 dwarfs with accurate parallaxes--only 13--this conclusion is tentative. However, it suggests a need to revisit existing brown  dwarf absolute magnitude relations with a larger sample of bona fide single L and T dwarfs with accurate spectrophotometry and parallaxes.

\section{Summary and Conclusions}
\label{sec:conclusion}

We have presented a 5.4$-$14.0 $\micron$ $R\sim100$ high-SNR (90--1000) spectral library of brown dwarfs across the L0--T6 spectral type range, as observed with \textit{JWST} MIRI LRS. Using this spectral library, we have produced the following results. 

\begin{itemize}
\item We have confirmed the onset of silicate absorption no later than spectral type L0, with varying degrees of silicate absorption between the L0 dwarfs in our observed sample.
\item We produced the first mid-IR detection of molecular SiO in L0--L2.5 dwarfs and further trace it to M dwarfs as early as M5.5 in archival \textit{Spitzer} IRS spectra. We found indications of gravity sensitivity for this absorption, with lower gravity L0--L2.5 dwarfs showing reduced SiO absorption. This is contrary to the behaviour at spectral type M, where mid-IR SiO is promiment in M giants, while we find its signature to be more tenuous in archival \textit{Spitzer} IRS spectra of M dwarfs.
\item We report the possible detection of CS$_2$ in a T dwarf, mirroring previous detections in solar system or extrasolar planets. This would be the first detection CS$_2$ in a substellar atmosphere outside a planetary system. Its confirmation would require higher-dispersion spectroscopy. 
\item We have newly resolved two binaries using our target acquisition imaging, confirmed two unresolved candidate near-equal flux binaries from their consistent near-IR through mid-IR overluminosity, and found additional evidence for two unresolved possible unequal-flux binaries.
\item Comparing our well-characterized observed sample to existing empirical relations, we found evidence for contamination of the current field sample by unresolved binaries, that lead to a 0.15--0.20 mag over-estimation of absolute magnitudes across the NIR and mid-IR.
\end{itemize}

Our spectral library will be useful as a source for several future fields of study. Dedicated modelling of silicate condensate compositions and properties, including the use of retrievals \citep{Burningham2021, Vos2023, Molliere2025} will enable a deeper analysis than ever before of the nature of silicate condensates in mid-L dwarfs. These spectra may also serve as templates for comparison with upcoming observations of other brown dwarfs, highlighting smaller discrepancies than previously possible such that peculiar dwarfs may be better identified and studied. 

Potential applications to exoplanet spectroscopy also exist. Directly imaged exoplanets appear spectroscopically similarly to low-gravity brown dwarfs \citep[e.g.,][]{Faherty2016}. Unlike young exoplanets, isolated brown dwarfs are not impacted by contrast issues with host stars, thus spectra of these dwarfs can be more easily obtained than those of exoplanets. These spectra can then be used as benchmarks when compared with directly imaged exoplanets, giving this spectral library utility even outside the brown dwarf field.

\section*{Acknowledgements}
We thank Greg Sloan for his advice on potential origins of the $\sim$7.5--9.5\,$\umu$m feature in our early L dwarfs. 
We acknowledge the support of the Canadian Space Agency (CSA) through grant agreement [23JWGO2A12]. PAMP acknowledges support from the grants RYC2021-031173-I and PID2022-137241NB-C42, funded by MCIN/AEI/10.13039/501100011033 and by the European Union NextGenerationEU/PRTR.
JMV acknowledges from a Royal Society - Research Ireland University Research Fellowship (URF/1/221932, RF/ERE/221108) and the European Union through the Exo-PEA ERC project (grant number 101164652). Views and opinions expressed are however those of the author(s) only and do not necessarily reflect those of the European Union or the European Research Council Executive Agency. Neither the European Union nor the granting authority can be held responsible for them.

This research made use of \textsc{Photutils}, an \textsc{Astropy} package for
detection and photometry of astronomical sources \citep{Bradley2025}.
This work is based on observations made with the NASA/ESA/CSA James Webb Space Telescope. The data were obtained from the Mikulski Archive for Space Telescopes at the Space Telescope Science Institute, which is operated by the Association of Universities for Research in Astronomy, Inc., under NASA contract NAS 5-03127 for \textit{JWST}. These observations are associated with program \#3930.

\section*{Data Availability}

The \textit{JWST} data analysed in this paper are available on MAST with program ID GO 3930.



\bibliographystyle{mnras}
\bibliography{refs} 








\bsp	
\label{lastpage}
\end{document}